\documentstyle[12pt,a4]{article}
\title{Superheavy Nuclei in the Relativistic Mean Field Theory}
\author{G.A. Lalazissis$^{1,4}$, M.M. Sharma$^2$, P. Ring$^1$ and 
Y.K. Gambhir$^3$\\
\\
$^1$Physik Department, Technische Universit\"at M\"unchen \\
D-85747 Garching, Germany\\
$^2$Max Planck Institut f\"ur Astrophysik,\\
D-85740 Garching bei M\"unchen, Germany\\
$^3$Physics Department, I.I.T. Powai, Bombay 400076, India\\
$^4$Department of Theoretical Physics, Aristotle University of Thessaloniki,\\
GR 54006 Thessaloniki, Greece.}

\begin{document}
\maketitle
\begin{abstract}
\baselineskip=17pt
We have carried out a study of superheavy nuclei in the
framework of the Relativistic Mean-Field theory.
Relativistic Hartree-Bogoliubov (RHB) calculations have
been performed for nuclei with large proton and neutron
numbers. A finite-range pairing force of Gogny type has
been used in the RHB calculations. The ground-state
properties of very heavy nuclei with atomic numbers
$Z$=100-114 and neutron numbers $N$=154-190 have been
obtained. The results show that in
addition to $N$=184 the neutron numbers $N$=160 and $N$=166
exhibit an extra stability as compared to their neighbors.
For the case of protons the atomic number $Z$=106 is shown
to demonstrate a closed-shell behavior in the region of
well deformed nuclei about $N$=160.  The proton number
$Z$=114 also indicates a shell closure. Indications for a
doubly magic character at $Z$=106 and $N$=160 are observed.
Implications of shell closures on a possible synthesis of
superheavy nuclei are discussed.
\end{abstract}
\newpage

\baselineskip=22pt
\section{\sf Introduction}

Exploration of the domain of superheavy nuclei has been
pursued for a long time and limits on stability and
feasibility of creating superheavy nuclei have been tested
time and again.  This pursuit has been enlivened by the
constant hope of creating nuclei having mass and charge
much larger than those we are familiar with. The shell
effects which play a major role in creating nuclei with
magic numbers and thus provide a higher stability have
increased the hope of being able to create superheavy
nuclei.  Theoretically, various schemes have been adopted
to calculate shell effects in the unknown teritorry of
superheavies \cite{Nix.77,Sob.94,MN.94} and at the same time
superheavy nuclei have evoked an enormous experimental
interest.  Recent discoveries of several new elements and
the ability of the experimentalists to synthesize heavy
nuclei with atomic numbers $Z$=109-112 have added to the
momentum of the activity in the pursuit of the superheavy
nuclei \cite{Amb.85,OL.85,Mun.88,Amb.94,Oga.95,Hof.95,Hof.96}.
However, an increase in the charge of a nucleus by every
unit renders the nuclei increasingly unstable and
consequently the ensuing nuclei live for a far shorter time
than those with a correspondingly lower charge.
Notwithstanding enormous difficulties in the synthesis and
detection of these highly rare entities, experimental
efforts are being made currently at the laboratories in
GSI, GANIL, Berkeley and Dubna.

Recently, the production and decay of the heavy element
$^{269}$112 have been reported \cite{Hof.96} by the GSI
group.  Using a different experimental setup the
Dubna-Livermore collaboration has discovered \cite
{Laz.94,Laz.95,Laz.96} new isotopes $^{265}$106, $^{266}$106
and $^{273}$110 and has measured their decay properties. It
has been inferred that there is an enhanced nuclear
stability in the vicinity of the deformed shells $N$=162
and $Z$=108. Nuclei in this region have been predicted to
be extra stable by some theories.  Accordingly experimental
efforts are currently being devoted to explore the region
about $Z$=108-116.

Theoretically, it is expected that a magic proton number
should exist at $Z$=114. These predictions are based
primarily upon phenomenological models such as finite-range
droplet model (FRDM) \cite{MN.94}. In addition to predicting
major shell gaps at $Z$=114 and $N$=184, the FRDM also
predicts larger shell gaps at proton numbers $Z$=104, 106,
108 and 110 at neutron numbers $N$=162 and 164. A prolate
deformation in this region has been surmised for these
nuclei \cite{MN.94}. For a search of superheavy nuclei,
theories based upon Nilsson-Strutinsky scheme
\cite{pat91,pat91b,Sob.94} have also been employed
extensively.  A similar pattern of deformed nuclei have
been predicted about $Z$=108 and $N$=162 in this approach
as in FRDM. In addition, microscopic calculations
\cite{bei74,quent78,GRT.90,Dob.96} have also been attempted
in this region. However, the main obstacle which the
theories including those of the macroscopic nature face is
the question whether the approaches which apply to the
region of beta-stability line can be extrapolated to such
very heavy systems.

Shell effects play a key role for the very existence of
magic nuclei. Such shell effects manifest strongly along
the line of stability in the form of much higher stability
of magic nuclei and thus a higher abundance of such
elements as compared to their neighbors. A semblance of the
same would be affected also for superheavy nuclei, if there
were any magic numbers in this region. Consequently, these
nuclei will be guarded against a faster decay by fission as
compared to their non-magic counterparts. Synthesis of
superheavy nuclei is thus subject to the interplay of the
shell effects in the region of very heavy nuclei.

The Relativistic Mean-Field (RMF) theory has recently
proven to be a very powerful tool for an effective
microscopic description of the ground-state properties of
nuclei \cite{GRT.90,rein89,SNR.93,LS.95,LSR.95,SLH.94}.  The RMF
theory has also been successful in describing the
properties of nuclei which entail shell effects.  Examples
where shell effects play an important role and where the
RMF theory has shown a remarkable success are the
description of the anomalous isotope shifts of many nuclei
from $Z$=40 region \cite{LS.95} to the rare-earth nuclei
\cite{LSR.95} and the most notable case of Pb nuclei
\cite{SLR.93}. A description of the deformation properties
and the complex series of shape transitions in many
isotopic chains has also been achieved \cite{LSR.95}, where
results including those with very large isospins match the
predictions of the exhaustive mass models FRDM and ETF-SI
(Extended Thomas-Fermi and Strutinsky Integral).  Thus, the
RMF theory has achieved a great success in providing a
unified description of the binding energies and deformation
properties of nuclei all over the periodic table including
exotic nuclei.  This gives an added confidence in the RMF
theory to employ and extrapolate it in the region of the
superheavy nuclei. It is noteworthy that the RMF theory
with scalar self-coupling employs only 6 parameters. This
is in contrast with the macroscopic-microscopic approaches
which use a considerably large number of parameters fitted
extensively to a large body of nuclear data.

Usually pairing correlations are taken into account only in
a very phenomenological way using occupation numbers of the
BCS type based on empirical pairing gaps deduced from
odd-even mass differences.  This procedure works well in
the valley of beta-stability, where experimental masses are
known. The predictive power for pairing gaps for nuclei far
from the line of beta stability and for superheavy nuclei
is thus limited.  We have, therefore, extended relativistic
mean-field theory to relativistic Hartree-Bogoliubov theory
\cite{KR.91} whereby we use a pairing force of finite
range, similar to that of the well established Gogny type
in non-relativistic calculations.

Using this pairing interaction we investigate the possible
existence of proton (neutron) shell closures in the region
of the super heavy nuclei, where it is well known that the
role of pairing is important. In our calculations we have
adopted the RMF force NL-SH \cite{SNR.93} which is
especially suitable for exotic nuclei as it takes into
account the isospin dependence correctly.

In the present paper, we investigate the ground-state
properties of nuclei in the region of a plausible existence
of superheavy nuclei. We have undertaken extensive
calculations for nuclei over a large range of atomic charge
and mass. In section 2 we describe some essential features
of the RMF theory. We give numerical and other details in
section 3. In section 4 results of the RMF calculations are
provided and discussed. A comparison of our results is made
with the predictions of other models wherever possible.  In
the last section we discuss our results vis-a-vis
experimental data available to-date and summarize our main
conclusions.

\section{\sf The Relativistic Mean-Field Theory}

The basic Ansatz of the RMF theory is a Lagrangian density
\cite{SW.86,Ser.92} whereby nucleons are described as Dirac
particles which interact via the exchange of various
mesons. The Lagrangian density can be written in the form:
\begin{equation}
\begin{array}{rl}
{\cal L} &=
\bar \psi (i\rlap{/}\partial -M) \psi +
\,{1\over2}\partial_\mu\sigma\partial^\mu\sigma-U(\sigma)
-{1\over4}\Omega_{\mu\nu}\Omega^{\mu\nu}+\\
\                                        \\
\ & {1\over2}m_\omega^2\omega_\mu\omega^\mu
-{1\over4}{\vec R}_{\mu\nu}{\vec R}^{\mu\nu} +
 {1\over2}m_{\rho}^{2} \vec\rho_\mu\vec\rho^\mu
-{1\over4}F_{\mu\nu}F^{\mu\nu} \\
\                              \\
 &  g_{\sigma}\bar\psi \sigma \psi~
     -~g_{\omega}\bar\psi \rlap{/}\omega \psi~
     -~g_{\rho}  \bar\psi 
      \rlap{/}\vec\rho
      \vec\tau \psi
     -~e \bar\psi \rlap{/}A \psi
\end{array}
\end{equation}
The meson fields included are the isoscalar $\sigma$ meson,
the isoscalar-vector $\omega$ meson and the
isovector-vector $\rho$ meson. The latter provides the
necessary isospin asymmetry. A correct isovector coupling
constant is important for describing the properties of
nuclei over a large range of isospins. The arrows in Eq.
(1) denote the isovector quantities.  The Lagrangian
contains also a non-linear scalar self-interaction of the
$\sigma$ meson :
\begin{equation}
U(\sigma)~={1\over2}m_{\sigma}^{2} \sigma^{2}~+~
{1\over3}g_{2}\sigma^{3}~+~{1\over4}g_{3}\sigma^{4}
\end{equation} 
The scalar potential (2) is essential for appropriate
description of surface properties \cite{BB.77}.  M,
m$_{\sigma}$, m$_{\omega}$ and m$_{\rho}$ are the nucleon-,
the $\sigma$-, the $\omega$- and the $\rho$-meson masses
respectively, while g$_{\sigma}$, g$_{\omega}$, g$_{\rho}$
and e$^2$/4$\pi$ = 1/137 are the corresponding coupling
constants for the mesons and the photon.

The field tensors of the vector mesons and of the
electromagnetic field take the following form:
\begin{eqnarray}
\Omega^{\mu\nu} =& \partial^{\mu}\omega^{\nu}-\partial^{\nu}\omega^{\mu}\\
\          \\
\vec R^{\mu\nu} =& \partial^{\mu}\vec\rho^\nu
                  -\partial^{\nu}\vec\rho^\mu\\
\                \\
F^{\mu\nu} =& \partial^{\mu}A^{\nu}-\partial^{\nu}A^{\mu}
\end{eqnarray}
The variational principle gives the equations of motion.
In our approach, where the time reversal and charge
conservation is considered, the Dirac equation for the
static case is written as:
\begin{equation}
\{ -i{\mbox \boldmath \alpha}{\mbox \boldmath \nabla} + 
V({\bf r}) + \beta [ M +S({\bf r}) ] \} 
\psi_i~=~\varepsilon_i\psi_i,
\end{equation}
where $V({\bf r})$ represents the vector potential:
\begin{equation}
V({\bf r}) = g_{\omega} \omega_{0}({\bf r}) + g_{\rho}\tau_{3} {\bf {\rho}}
_{0}({\bf r}) + e{1+\tau_{3} \over 2} A_{0}({\bf r}),
\end{equation}
and $S({\bf r})$ is the $scalar$ potential:
\begin{equation}
S({\bf r}) = g_{\sigma} \sigma({\bf r}) 
\end{equation}
the latter contributes to the effective mass as:
\begin{equation}
M^{\ast}({\bf r}) = M + S({\bf r}).
\end{equation}
The Klein-Gordon equations for the meson fields are
time-independent inhomogeneous equations with the nucleon
densities as sources.
\begin{eqnarray}
\{ -\Delta + m_{\sigma}^{2} \}\sigma({\bf r})
&=& -g_{\sigma}\rho_{s}({\bf r})
-g_{2}\sigma^{2}({\bf r})-g_{3}\sigma^{3}({\bf r})\\
\{ -\Delta + m_{\omega}^{2} \} \omega_{0}({\bf r})
&=& g_{\omega}\rho_{v}({\bf r})\\
\{ -\Delta + m_{\rho}^{2} \}\rho_{0}({\bf r})
&=& g_{\rho} \rho_{3}({\bf r})\\
 -\Delta A_{0}({\bf r}) &=& e\rho_{c}({\bf r})
\end{eqnarray}
The corresponding source densities are obtained as
\begin{equation}
\begin{array}{ll}
\rho_{s} =& \sum\limits_{i=1}^{A} \bar\psi_{i}~\psi_{i}\\
\             \\
\rho_{v} =& \sum\limits_{i=1}^{A} \psi^{+}_{i}~\psi_{i}\\
\             \\
\rho_{3} =& \sum\limits_{p=1}^{Z}\psi^{+}_{p}~\psi_{p}~-~
\sum\limits_{n=1}^{N} \psi^{+}_{n}~\psi_{n}\\
\                    \\
\ \rho_{c} =& \sum\limits_{p=1}^{Z} \psi^{+}_{p}~\psi_{p}
\end{array}
\end{equation}
where the sums are taken over the valence nucleons only.
In the present approach contributions from negative-energy
states are neglected ($no$-$sea$ approximation), i.e. the
vacuum is not polarized. Thus, we work within the framework
of the non-linear $\sigma \omega$ model which includes the
scalar self-coupling up to quartic order.

\section{\sf Details of the Calculations}

Calculations for superheavy nuclei have been performed
using the oscillator expansion method. The details of the
method have been provided in our earlier papers (for
example see Ref. \cite{GRT.90}). Most of our earlier
calculations were carried out using 20 oscillator shells
for spherical nuclei and 12 or 14 oscillator shells for
deformed nuclei.  In the present case of very heavy nuclei
we have used 20 shell for spherical as well as for deformed
nuclei.

It is well known that pairing correlations play an
important role in our understanding of structure of nuclei
with open shells.  Using Green's function techniques and
Gorkov's factorization one can derive in principle a
relativistic theory of pairing correlations \cite{KR.91}:
Starting from a Lagrangian containing nucleonic spinors and
meson fields one obtains relativistic Hartree-Bogoliubov
equations.  If one uses the same parameters for the mesons
in the particle-particle channel as in the relativistic
mean-field theory, a quantitative description of pairing
is, however, not possible. The large values of the
$\sigma$- and $\omega$-mass lead to relatively short-range
interactions which produce much too strong pairing
correlations. Thus, in the relativistic theories one is
faced with the same problem as with Skyrme forces, i.e. one
needs cut off parameters in the particle-particle channel.
In principle, however, there is no reason why one should
use the same effective interaction in both the channels.
At present there is no microscopic derivation of the
effective interaction responsible for particle-particle
correlations. However, in many cases the simple monopole
pairing force with constant pairing matrix elements in a
certain region around the Fermi surface turns out to yield
very satisfactory results, if the pairing force constant
$G$ is adjusted to the experimental gap parameter deduced
from the odd-even mass differences.  This prescription is
used in many ways, for instance, in Nilsson-Strutinsky
calculations, in density-dependent Hartree-Fock
calculations with Skyrme forces and also in the
relativistic mean-field theory. Thus, in this procedure one
ends up with rather different pairing force constants $G$
in the different regions of the periodic table. In fact,
the size of this constant depends on the pairing cut off.
On the other hand, Gogny's parameterization of an effective
force based on two Gaussians with a finite range provides a
unified phenomenological description without a cut-off
parameter of pairing properties across a large part of the
periodic table\cite{DG.80}.  This method is considered to
work satisfactorily in the region of medium and heavy
nuclei.
   
In this paper, we therefore use in the particle-particle
channel a finite-range force of the Gogny type This force
consists of a sum of two Gaussians,
\begin{equation}
V^{pp}(1,2)~=~\sum_{i=1,2} 
e^{-\left(\frac{\mbox{\boldmath $r$}_1-\mbox{\boldmath $r$}_2} 
{\mu_i}\right)^2}\,
\left(W_i~+~B_i P^\sigma -H_i P^\tau -
M_iP^\sigma P^\tau\right), 
\end{equation}
with the parameters $\mu_i$, $W_i$, $B_i$, $H_i$, and $M_i$
$(i=1,2)$.  a short-range repulsive term and a medium-range
attractive term.  We neglect the density-dependent part and
the spin-orbit part of the Gogny force, because the
density-dependent part vanishes in the $S=0$, $T=1$ channel
and the latter gives only relatively small contributions.
Since we are working in an oscillator basis, we can use the
very elegant and simple techniques introduced by Talman
\cite{Tal} in order to evaluate the Gaussian matrix
elements in a spherical basis. In this paper, we use the
parameter set D1 for the pairing force as given in Ref.
\cite{DG.80} and are listed in Table 1.

Calculations have also been performed for a set of nuclei
in a cylindrically symmetric deformed configuration. So far
we have not implemented the evaluation of the Gogny matrix
elements in the axially deformed bases. We therefore use
for the deformed calculations the usual BCS formalism with
constant pairing gaps obtained from the prescription of
Ref. \cite{MN.92}.  In these calculations also we have now
included 20 oscillator shells both for the fermionic as
well as bosonic wave functions. The calculations have been
carried out only for a selected set of nuclei in view of an
enormously large computation time required for a Hartree
minimization in a deformed basis with 20 shells.

In our study the force NL-SH \cite{SNR.93} has been used.
The parameters of NL-SH are given in Table 2.  It has been
shown that NL-SH reproduces a wide variety of nuclear data
all over the periodic table. The shell effects and an
appropriate symmetry energy contained in this force are
responsible for describing various data successfully.

\section {\sf Results and Discussion}

\subsection{\sf Particle Separation Energies}

The present study of superheavy nuclei spans even-even
nuclei with atomic numbers $100 \leq Z \leq 118$ and
neutron numbers $154 \leq N \leq 190$. First, we present
the results of the relativistic Hartree-Bogoliubov (RHB)
calculations for nuclei with $158 \leq N \leq 190$ assuming
a spherical configuration and employing the force NL-SH.
In these calculations the finite-range pairing force of the
Gogny type has been used for pairing as discussed above.
This minimizes the uncertainties in the pairing

For the sake of a clear presentation of the results the
neutron range 158-190 has been divided into two parts,
158-176 and 178-190.  Fig. 1 (a) shows the two-neutron
separation energies $S_{2n}$ for nuclei with $N$=158-176
for the whole range of the atomic number $Z$=100-114. Each
curve corresponds to an isotopic chain for a given $Z$. The
lowest curve is for $Z$=100 and the highest one corresponds
to $Z$=114. Clearly, with an increase in the neutron
number, the $S_{2n}$ values show a regular decrease except
at $N$=164-166, where a slight discontinuity in the
two-neutron separation energies can be seen. This
discontinuity is more pronounced above $Z$=110.  Such
discontinuities are symptomatic of shell effects which
prevail all over the periodic table. Similar effects emerge
also in the compilation of recent empirical masses
in Ref. \cite{AW.93}, where clear discontinuities
at the known magic numbers can be seen. In the present
case, nuclei above $N$=164 become increasingly vulnerable
to neutron decay due to a fast decrease in the $S_{2n}$
values.  This is indicative of an enhanced stability for
nuclei with $N$=164. The curve for $Z$=106, in contrast,
shows an increase in $S_{2n}$ value at $N$=166.  This
indicates that the nucleus with $N$=166 and $Z$=106 is
slightly more stable against neutron decay.

The two-neutron separation energies in the RMF theory for
the second range of neutrons $N$=178-190 are shown in Fig.
1 (b).  The curves for the range of the proton numbers
$Z$=100-114 are shown. The lowest curve corresponds to
$Z$=100 and the highest one is for $Z$=114.  A strong kink
in the $S_{2n}$ values is clearly visible for the curves
$Z$=100-108 at $N$=184. The kink for the other $Z$ values
decreases slightly and it becomes relatively much smaller
at $Z$=114.  This kink at $N$=184 for all $Z$ values
underlines the manifestation of a magic number at $N$=184.
The shell-closure at $N$=184 is consistent with the
predictions of other theoretical models such as
Nilsson-Strutinsky and finite-range droplet model (FRDM)
\cite{MNM.95}, whereby a strong shell-closure at this
neutron number has been suggested.

The two-proton separation energies $S_{2p}$ obtained in the
RMF theory for nuclei with $Z$=102-114 are shown in Fig. 2
(a). Each curve corresponds to a given neutron number which
changes from $N$=156 to $N$=178 in going from the bottom to
the top of the figure. A decreasing trend with an increase
in the atomic number is to be seen clearly. The $S_{2p}$
values also show an obvious decrease in the values with a
decrease in the neutron number. However, a small kink in
the $S_{2p}$ values at $Z$=106 for the neutron numbers
$N$=156-160 is observed. This kink is reduced as one
proceeds to neutron numbers higher than $N$=156 and it
vanishes for nuclei above $N$=160. Thus, spherical
calculations for these nuclei demonstrate the existence of
a large shell gap at $Z$=106 with neutron numbers
$N$=156-160. This shell gap in the proton number is washed
out in going to higher neutron numbers.  In section 4.6 on
shell corrections, it will be shown that the shell
correction energies do corroborate to a magic like
character for $Z$=106 for neutron numbers about $N$=160.

The corresponding $S_{2p}$ values for nuclei with higher
neutron numbers $N$=180-188 are shown in Fig. 2 (b). The
values show a monotonous decrease with an increase in
proton number. There is no kink to be seen in the $S_{2p}$
values. Thus, for nuclei with even higher neutron numbers
than those shown in Fig. 2 (a), the spherical calculations
do not show any magic proton number except for $Z$=114,
where a slight change in the slope is indicated.

For the sake of a qualitative comparison, results obtained
on the two-neutron separation energies in the FRDM are
shown in Figs. 3 (a)-(b). A kink about $N$=162 and $N$=184
is seen clearly in both the figures. A similar feature was
also predicted in density-dependent Skyrme theory
\cite{bei74}.  On the conclusion that $N$=184 is a magic
number, various theories including the RMF theory as well
as other non-relativistic approaches are unanimous. In
comparison, the mass model Extended Thomas-Fermi with
Strutinsky Integral (ETF-SI) \cite{APD.92}, which is based
upon the Skyrme Ansatz, does not exhibit any clear
signature for the magicity about $N$=162. However, it does
show a slight kink in the $S_{2n}$ values at about $N$=184,
as can be seen in Figs. 4 (a) and (b).

The scenario for the mass models FRDM and ETF-SI is
different as far as $S_{2p}$ values are concerned. In the
FRDM, the $S_{2p}$ values show hardly any kink about
$Z$=106 (Fig. 3(c)). Probably it is because the FRDM
predicts larger shell gaps for the proton numbers at
$Z$=104, 106, 108 and 110 consecutively and as a result
indications of a discontinuity amid these proton numbers
seem to disappear.  This is, however, different with the
RMF predictions, where evidence for a 'magic' proton number
at $Z$=106 was obtained for neutron numbers in the region
of $N$=160.  The FRDM values, on the other hand, show a
clear kink only at $Z$=114 for higher neutron numbers about
$N$=184 (Fig. 3(d)). The ETF-SI values, in contrast with
FRDM, show a clear kink at $Z$=106 for lower neutron
numbers $N$=156-166 (Fig. 4(c)).  There is, however, no
kink around $Z$=114 for higher neutron number such as
$N$=184 (Fig. 4(d)).  This observation in the ETF-SI is at odds
with most of the theories which predict a strong magic
number at $Z$=114.  However, it should be noted that the recent
density dependent Hartree-Fock-Bogoliubov calculations with
the Skyrme force SkP \cite{Dob.96} also does not predict a
magic number at $Z$=114.

\subsection{\sf Pairing Energy}

Pairing energy provides a reliable indication of magicity
of a particle number. For magic nuclei, single-particle
levels up to Fermi energy are fully occupied and hence
there is no smearing of the Fermi surface. This implies
that in such cases pairing is non-existent and hence the
pairing energy should vanish. This feature is usually
reflected in the sequence of the single-particle levels
followed by a large gap in the levels near the Fermi
energy.

In our self-consistent relativistic Hartree-Bogoliubov
calculations, we have used finite-range pairing of the
Gogny type, whereby the particle levels and the pairing
fields are calculated self-consistently. The corresponding
pairing energies for neutrons and protons as obtained in
the RHB calculations are shown in Figs. 5 and 6,
respectively.

The neutron pairing energies for several chains of nuclides
with $Z$=100 to $Z$=114 are given in Fig. 5. A very strong
peak in the pairing energy is observed at $N$=184 for all
the atomic numbers. The pairing energy for all the nuclei
with $N$=184 is seen to be zero. This fact is in accord
with a strong kink observed in the $S_{2n}$ values at
$N$=184 as shown in Fig. 1 (b).  Thus, it is demonstrated
that the neutron number $N$=184 constitutes a very strong
magic number. In addition, we also observe two other peaks
in the neutron pairing energy, albeit not so strong, at
$N$=164 and $N$=172. These peaks are highest for $Z$=114.
The peak structure at the above neutron numbers diminishes
gradually in going down from $Z$=114 to $Z$=100.
Comparatively, the peak at $N$=164 still remains up until
$Z$=100, whereas the peak at $N$=172 disappears fast in
going down to lower atomic numbers. As regards to the
magnitude of the pairing energy for $N$=164 and $N$=172, it
is seen that the pairing energy does not vanish even for
$Z$=114 whereby the peak structure is most prominent
amongst various atomic numbers.  The lowest pairing energy
at $N$=164 is about $-1.0$ MeV for the $Z$=114 nuclide.
This value of the pairing energy is very small and
indicates that the neutron pairing in this case is minimal.
Thus, the results of Fig. 5 provide a good indication for
the neutron number $N$=164 showing a relatively strong
magicity.

The corresponding proton pairing energy is shown for
nuclear chains from $N$=156 to $N$=190 in Fig. 6. A very
strong peak at $Z$=106 is observed. The proton pairing
energy vanishes at $Z$=106 for most of the nuclides with
$N$=156 to $N$=164.  The pairing energy for nuclei with
$Z$=104 or $Z$=108 is 4-6 MeV larger and the pairing energy
increases even further in going away from $Z$=106. This
emphasizes the predominance of the $Z$=106 peak in the
proton pairing energy.  Thus, $Z$=106 turns out to possess
a strong magic character.  It is noteworthy that for nuclei
with neutron numbers higher than $N$=164, the pairing
energy is no longer zero and the peak at $Z$=106 vanishes.
Thus, $Z$=106 does not show a magic character for higher
neutron numbers. In conjunction with the neutron pairing
energies of Fig. 5, we find that the nucleus with $Z$=106
and $N$=164 behaves like a doubly magic nucleus. An
additional small peak is observed at $Z$=114 indicating a
small magicity at this proton number.

\subsection{\sf Alpha-decay Half-Lives}

Alpha decay is one of the most predominant modes of decay
of superheavy nuclei. Depending upon the region of an extra
stability which would originate from shell gaps and
magicity, the half-life of the alpha decay is another
indicator about a possible valley of stability. For an area
of enhanced stability, the alpha-decay half-lives are
expected to be longer than its neighbors. With this view,
we calculate the alpha-decay half-lives for several
isotopic chains. We employ the phenomenological formula of
Viola and Seaborg \cite{viola} for calculation of $\alpha$
half-lives:

\begin{equation}
\log T_{\alpha} = (aZ + b) Q_{\alpha}^{-1/2} + (cZ +d )
\end{equation}
where $Z$ is the atomic number of the parent nucleus and
Q$_{\alpha}$ is the alpha-decay energy in MeV. T$_{\alpha}$
is then given in seconds. The parameters $a$, $b$, $c$, and
$d$ are taken from Ref \cite{sobi89}, where these have been
readjusted in order to take into account new data. The
explicit values for these parameters are: $a$=1.66175, 
$b$=-8.5166, $c$=-0.20228, and $d$=-33.9069.

In Fig. 7 we plot the half lives $T_{\alpha}$ using the
Viola and Seaborg \cite{viola} systematics. The logarithm
of the half lives is shown for each isotopic chain in the
region $Z$=102-118. One observes an enhancement in the
$\log T_\alpha$ values at $N$=164 followed by a decrease at
$N$=166 for most of the chains. The peak at $N$=164 emerges
clearly in going to higher atomic numbers above $Z$=108.
Thus, the neutron number $N$=164 would support synthesis of
superheavy nuclei above $Z$=108.

Another maximum in the half-lives is seen at $N$=184.  This
is accompanied by a strong plateau up until $N$=184 for
nuclei with lower atomic numbers. This plateau is followed
by a strong dip at $N$=186 for nuclei below $Z$=112. The
sudden and drastic change in the alpha-decay half-lives at
$N$=186 is a strong indication of the magicity of $N$=184.
Thus, the $T_\alpha$ values indicate regions of extra
stability in the vicinity of $N$=164 and $N$=184, whereby
the magicity of $N$=184 has been demonstrated
unambiguously.

It should be noted that the half-lives $T_\alpha$ shown in
Fig. 7 are obtained from the spherical RHB calculations.
However, as we will see in the next section, some of these
nuclei are deformed and therefore some of the
$Q_\alpha$-values may change slightly. Even a slight change
in the $Q_\alpha$'s may produce half-lives $T_\alpha$
different by orders of magnitude.  Therefore Fig. 7 should
be regarded only as an indicative of the general trend.

\subsection{\sf Calculations with Deformed Configurations}
 
In most of our previous investigations, calculations for
deformed nuclei have been limited to a maximum of 14
deformed oscillator shells in the expansion, constrained
mainly by computational reasons.  For nuclei not so heavy
in charge and mass, calculations with 12 or 14 shells do
produce reliable results. An extension of the deformed RMF
calculations by taking into account higher number of
oscillator shells up to 20 is extremely time consuming.
Therefore, deformed calculations for the whole of the
superheavy regions with 20 oscillator shells put a heavy
burden on computation.  In view of this, we have selected a
region of superheavy nuclei about the neutron number
$N$=164 and another one about $N$=184 for axially deformed
RMF calculations with 20 shells.  Pairing correlations are
included in the constant gap approximation.  The pairing
gaps used in the calculations were obtained from the
prescription of Ref. \cite{MN.92}.

The region with $N$=164 is likely to yield nuclei with a
reasonable deformation, whereas the latter region with
$N$=184 being a strong magic number is likely to provide
nuclei with an almost spherical shape. Results of the
axially deformed RMF calculations are shown in Table 3,
where the quadrupole and hexadecupole deformations
$\beta_2$ and $\beta_4$, respectively, are shown for
several isotopic chains. The $\beta_2$ and $\beta_4$ values
are obtained from the quadrupole and hexadecupole moments
using the method of Ref. \cite{LQ.82}.

The nuclei given in Table 3 have been selected with a view
to be able to calculate alpha-decay half-life of nuclei
around $N$=164. The calculations encompass a few isotopes
of nuclei with $Z$=102 and $Z$=110-112, whereas
calculations for many more isotopes have been performed for
nuclei with other atomic numbers.  Predictions on the
quadrupole and hexadecupole deformations in the FRDM and
ETF-SI mass models are also shown for comparison. It is
observed that most of the nuclei in this region are well
deformed. The $\beta_2$ values obtained in the RMF theory
are close to those of FRDM for most of the cases and both
the RMF theory and FRDM show a similar trend as a function
of mass number.  On the other hand, the ETF-SI model seems
to be using a rather fixed value of $\beta_2$ for most of
the nuclei in a given chain.  Therefore, a comparison of
the RMF values with the ETF-SI values is not meaningful. On
the whole, the RMF theory as well as the mass models, all
predict a prolate shape for all the nuclei in this region.

The $\beta_4$ values obtained in the RMF theory for all the
nuclei in Table 3 are negative. These values in the RMF
theory are comparable to those in the FRDM.  Moreover, the
RMF theory predicts the same sign (negative) for the
hexadecupole deformation as does the FRDM. The $\beta_4$
values in the ETF-SI are also negative for most of nuclei
with only a few exceptions for heavy nuclei with $A$ $>$
276. Thus, the RMF hexadecupole moments are by and large in
good conformity with the existing mass models.

The two-neutron separation energies ($S_{2n}$) and
alpha-decay half-lives ($T_\alpha$) obtained from the
deformed RMF calculations are shown in Fig. 8 for the
isotopic chains of $Z$=106 and $Z$=108. The $S_{2n}$ values
show a usual decrease with an increase in neutron number.
The $Z$=108 curve does not show any structure about
$N$=164, whereas $Z$=106 values do show a kink about
$N$=166. Thus, $S_{2n}$ values for $Z$=106 show an
indication of some magicity about $N$=164-166.  The
alpha-decay half-lives also display a clear structure above
$N$=164 and exhibit a significant enhancement in the
$T_\alpha$ value for nuclides with $N$=166 both for $Z$=106
as well as for $Z$=108. Thus, an extra stability is shown
by both the nuclei $^{272}$106 and $^{274}$108.  Nuclei
with $N$=168 show a slight decrease in the $T_\alpha$ value
as compared to nuclei with $N$=166. However, the $T_\alpha$
value for $N$=168 is considerably higher than for nuclei
with neutron numbers below $N$=166. Thus, these values
signify an area of an added stability about $N$=166. It is
to be remarked that the values of $T_\alpha$ shown in Fig.
8 differ from the corresponding values displayed in Fig. 7
due to slightly different values of $Q_\alpha$ obtained in
the deformed calculations.

Several recent experiments have been able to measure energy
of $\alpha$-particles emitted in decay of superheavy
nuclei. In order to facilitate a comparison of the
theoretical predictions with experimentally observed
alpha-decay energies, we show in Fig. 9, the Q-value for
alpha decay, i.e., $Q_\alpha$ for the isotopic chains with
$Z$=106 and $Z$=108.  The RMF theory predicts that the
isotopes with $Z$=106 would decay with an $\alpha$-particle
energy of about 7-8 MeV, whereas nuclei in the isotopic
chain with $Z$=108 would decay with an $\alpha$-particle
energy of about 9-10 MeV.  The measured energy of the
emitted $\alpha$ particle is 8.63 MeV for $^{266}106$ and
the corresponding energy for the case of $Z$=108 is between
9.7~-~9.87 MeV. These numbers are in close conformity with
the calculated $Q_\alpha$ shown in Fig. 9.  A further
comparison of these values with recent experimental
observations will be made in the last section.

In order to check the sphericity of nuclei in the vicinity
of $N$=184, we have performed RMF calculations in a
deformed basis also for several nuclei with $N$=184 by
taking into account 20 oscillator shells. The nuclei
considered are isotones of $N$=184 with mass numbers
$A$=290, 292, 294 and 298.  The binding energies and
quadrupole deformations obtained in the RMF calculations
are shown in Table 4.  The predictions of FRDM and ETF-SI
both on binding energies as well as on deformations are
also shown for comparison.  The RMF theory predicts the
quadrupole deformation of these nuclei to be very close to
zero. Thus, the RMF theory and mass models both predict a
spherical shape for nuclei with $N$=184. It may be
emphasized that moving away from the region of a strong
deformation about $N$=166 towards $N$=184, the RMF theory
produces nuclei which tend to become spherical as the
neutron number 184 is approached. This is again an
indication that the region about $N$=184 is associated with
a magic number.

The binding energies obtained in the RMF theory are in good
agreement with those of ETF-SI within 1-2 MeV, whereas the
FRDM values differ from the RMF and ETF-SI values by only a
few MeV. Thus, the RMF theory predicts binding energies
which are very close to those of FRDM and ETF-SI.

\subsection{\sf Single-Particle Spectra of Superheavy Nuclei}

The single-particle levels obtained in the deformed RMF
calculations for several key nuclei are shown in Figs.
10-13 both for neutrons and protons.  The levels correspond
to the ground-state deformations as obtained in each
calculation and which have been shown in Table 3. First, we
show the single-particle spectra for isotopic chains with
neutron numbers $N$=162-166 for the proton number $Z$=106
and $Z$=108 in Figs.  10-12. The numbers in the braces
shown in the large shell gaps denote shell closures. The
associated Fermi energies are also shown by dashed lines.

The single-particle (s.p.) spectrum for $^{268}$106
($N$=162) shown in Fig. 10 exhibits only a small gap at
$N$=162.  For protons a considerably larger gap at $Z$=106
in this nucleus can be seen. In the same figure, the
neutron s.p. spectrum for the nucleus $^{270}$108 ($N$=162)
does not display any significant gap near $N$=162. A clear
gap is, on the other hand, visible at the neutron number
$N$=166. Similarly, a major shell gap at proton number
$Z$=108 is seen. Thus, the single-particle spectra for the
above two isotones show a major shell gap at $N$=166 in
neutrons and an equally strong shell gap in protons for
both $Z$=106 and $Z$=108.

In Fig. 11 we examine the single-particle spectrum for
isotones with $N$=164. For nuclei with $Z$=106 as well as
$Z$=108, a clear gap in the neutron single-particle
spectrum emerges at $N$=166. This is consistent with what
we observed also in Fig. 10.  Similarly, a reasonably good
gap is to be seen for proton number $Z$=106 in the nucleus
$^{270}$106 as well as for $Z$=108 in the nucleus
$^{272}$108. In the single-particle spectrum for isotones
with $N$=166 (Fig. 12), a very clear shell gap at neutron
number $N$=166 can be seen. This gap is observed
consistently in the single-particle spectra of nuclei with
$N$=162, 164 and 166. This lends credence to the prediction
that a major shell gap at $N$=166 should exist.  This would
consequently provide a region of extra stability centered
about $N$=166.

The proton single-particle spectra for $N$=166 isotones
show a gap at proton number 106 both for the nuclei with
atomic numbers $Z$=106 as well as $Z$=108. Thus, the
nucleus with $N$=166 and $Z$=106 can be construed as a
'double magic' nucleus in the landscape of deformed nuclei
prevalent in this region.

Single-particle spectra for isotones of $N$=184 with
$Z$=106, 108, 110 and 114 are shown in Fig. 13 both for
neutrons and protons. A very clear and profound shell gap
at $N$=184 is exhibited by neutron single-particle spectra
for all the isotones. This is in conjunction with the fact
that nuclei around $N$=184 are predominantly spherical in
the RMF theory. This establishes the magicity of the
neutron number $N$=184 unequivocally. At the same time it
is observed that the shell gap at $N$=184 decreases
slightly as one proceeds from $Z$=106 to $Z$=114. This may
imply that though $N$=184 retains its magicity in going to
higher atomic numbers, the magicity does show a decline
towards higher atomic numbers. This may have  a consequence
that nuclei if synthesized about $N$=184 would tend to
favor a charge value $Z$=106 rather than $Z$=114. For the
latter a decreased stability stemming from a reduced shell
gap at $N$=184 will ensue.

The proton single-particle spectra for $N$=184 isotones
show a shell gap at $Z$=106 as well as at $Z$=114. It can
be seen that the gap at $Z$=106 is not so strong vis-a-vis
that observed for the major neutron shell-closure $N$=184.
The shell gap at $Z$=114 is, in comparison, as strong as
that exhibited by $N$=184. The gaps at $Z$=106 and at
$Z$=114 are strong indications of  shell closures in the
proton mean field. Thus, $Z$=106 as well as $Z$=114 show
signatures of magicity in the proton number. The nucleus
with $Z$=106 and $N$=184 and the one with $Z$=114 and
$N$=184  provide a semblance of doubly magic nuclei.

\subsection{\sf Shell Corrections}

Shell corrections provide an indicator about the deviation
in the structure of nuclei away from the smooth liquid-drop
type of behavior. The magnitude of the shell correction
signifies the role of shell effects at play in a nucleus.
We have evaluated shell correction for a set of nuclei
employing the Strutinsky procedure. The single-particle
spectra of nuclei as obtained in the RMF theory have been
used as an input for a smoothing procedure. Pairing has
been duly taken into account in these calculations
according to the prescription of Ref. \cite{BDJ.72}.

The microscopic shell corrections obtained from the
Strutinsky procedure applied to the single-particle spectra
of the RMF theory are given in Table 5 for several nuclei.
In these calculations we have covered many nuclei with
neutron numbers from $N$=154 to $N$=168 and with proton
numbers from $Z$=104 to $Z$=116. The table shows a
significant negative shell correction suggesting an
important role of the shell effects in this region.  We
scan the table row-wise. The first two rows, i.e. the shell
corrections for $Z$=104 and $Z$=106 show a large minimum at
-6.79 and -8.05 MeV, respectively. Both the minima
correspond to $N$=160. This implies that there should exist
a large shell gap at $N$=160. In the two-dimensional
landscape of the shell correction, the shell energy shows a
decrease from -6.79 MeV for $Z$=104 to -8.05 MeV for
$Z$=106. The magnitude of the shell energies for neutron
numbers less than $N$=160 in the first two rows is less
than those for $N$=160. The values, however, still indicate
a significant shell correction below $N$=160. For nuclei
above $N$=162, shell correction energies are much smaller
than for $N$=160.  Moreover, the minimum at $Z$=106 and
$N$=160 is the absolute minimum in the given table. Thus,
the nucleus $^{266}$106 with a large shell correction is
expected to exhibit a reasonably higher stability as
compared to its neighbors.  This scenario is consistent
with the experimental observation of the isotope
$^{266}$106 and with its decay properties \cite{Laz.94},
where an indication of an extra stability in the superheavy
nuclei has been conjectured.

Considering the shell corrections for nuclei above $Z$=106,
i.e. the rows below the first two, the minimum in the
energy shifts a slightly. For $Z$=108, a minimum occurs at
$N$=158, which is a reasonably strong one. The shell
correction energies for the next neighbors on both the
sides of $N$=158 are smaller only by 0.2 MeV. Thus, all the
nuclei with $N$=156, 158 and 160 and $Z$=108 are likely to
possess reasonable shell gaps in the single-particle
spectra and therefore an ensuing stability.

For nuclei above $Z$=108, the minimum branches off to
$N$=158 and $N$=166. As the proton number increases above
110, the minimum at $N$=158 diminishes gradually into a
local minimum and at the same time a stronger minimum
develops at $N$=166. Thus, the neutron number $N$=158 does
not benefit from the approaching $Z$=114 which is predicted
to be a major magic number. On the contrary, nuclei with
$N$=166 and $Z$=112, 114 could show a semblance of 'double
magic' character in the island of deformation. The nucleus
with $N$=166 and $Z$=112 has been shown to possess a
quadrupole deformation $\beta_2 = 0.18$ (Table 3 ). The
proton number $Z$=114, on the other hand, is expected to
form a doubly magic spherical nucleus with its neutron
counterpart at $N$=184, the region which is at present not
yet accessible experimentally.

\section{\sf Summary and Conclusions}

We have studied the ground-state properties of nuclei in
the superheavy region with a view to explore possible
regions of enhanced stability.  Calculations have been
carried out in the framework of the relativistic non-linear
$\sigma\omega$ model with scalar self-coupling. First, we
performed Hartree-Bogoliubov calculations using a spherical
configuration in an oscillator basis with 20 shells.  These
calculations employ the finite-range pairing force of the
Gogny type. It has been shown that the behavior of the
two-neutron separation energies and the pairing energy of
neutrons indicate a possible shell closure at $N$=164 and
$N$=184 in the neutron number. In the proton number, the
corresponding quantities suggest a possible shell closure
at $Z$=106 and $Z$=114.

Nuclei in the region of the neutron number $N$=166 are
expected to be deformed. Therefore, we have also performed
RMF calculations for nuclei on both the sides of $N$=166
using an axially deformed configuration. These calculations
use 20 oscillator shells, and the pairing has been included
in the BCS formalism using constant pairing gaps. The
calculations encompass atomic numbers from $Z$=102 to
$Z$=114. It has been shown that nuclei in the region of
$N$=166 acquire a reasonably strong quadrupole deformation
of the prolate type. Nearly all the nuclei in this region
exhibit a negative hexadecupole deformation. The RMF
results show a very good agreement with the predictions of
the mass formula FRDM both on the magnitude as well as on
the sign of the quadrupole and hexadecupole deformations.
The quadrupole and hexadecupole deformations in the RMF
theory can also be compared reasonably well with the
predictions of ETF-SI, where a constant value of $\beta_2$
and $\beta_4$ seems to  have been used in this region of
nuclei.

The half-lives of alpha-decay have been calculated from the
results of the deformed RMF calculations. The results
indicate a significant enhancement in the alpha-decay
half-life about $N$=166 and thus the RMF theory predicts a
region of an extra stability near $N$=166. This is
consistent with the results obtained in the spherical
calculations for the particle-separation energies. A
possible enhancement in the alpha-decay half-lives is also
observed at about $Z$=106-108 in the proton number.  It is,
however, difficult to say with certainty on the basis of
alpha-decay half-lives only, whether $Z$=106 or $Z$=108
constitutes a possible shell closure.

The single-particle spectra help to provide a clue as to
whether a particular particle number has a magic character
or not.  We have therefore examined the single-particle
spectra obtained from deformed RMF calculations. The
structure of the single-particle spectra reveals major
shell gaps in neutrons at $N$=166 and at $N$=184. It may be
recalled that in the spherical calculations a closed
neutron shell was obtained at $N$=164. Thus, the
deformation of nuclei in this region drives the closed
shell from $N$=164 to $N$=166. On the other hand, the shell
gap at the neutron number $N$=184 is shown unambiguously
also in the deformed calculations which lead to spherical
nuclei. It is consistent with the predictions of the
calculations in the spherical basis.  It may be appropriate
to say that the shell gap at $N$=166 would bear a strong
significance to creating superheavy nuclei, as it indicates
a 'semi-magic' nature for this neutron number.

The structure of the proton single-particle spectra, on the
other hand, provides a signature for a seemingly major
shell gap at $Z$=106. A shell gap at $Z$=108 is, however,
not observed in the single-particle spectra. Thus, in the
domain of protons, $Z$=106 is the only 'semi-magic' number
predicted by the RMF theory in the region of lighter
superheavy nuclei. This is slightly different from the
results of the Nilsson-Strutinsky approach \cite{Sob.94}
which predicts a strong shell-closure in protons at
$Z$=108.

It is noteworthy that the recent discovery of superheavy
nuclei $^{265}$106, $^{266}$106 \cite {Laz.94} and
$^{267}$108 \cite{Laz.95} and measurements of the associated
properties shed a considerable light on some of the issues
related to this unknown region. The Dubna-Livermore
collaboration has observed \cite{Laz.94} the alpha-decay
half-life of the above $Z$=106 isotopes to be 2-30 s and
10-30 s, respectively.  The corresponding alpha-decay
energies have also been measured and have been found to be
8.71-8.91 MeV and 8.63 MeV, respectively.  In comparison,
the half-life of the isotope $^{267}$108 has been measured
\cite{Laz.95} to be $19 ^{+20}_{-10}$ ms with an
$\alpha$-decay energy of 9.74-9.87 MeV. The half-life for
the $Z$=108 isotope is 3 orders of magnitude smaller than
those of the $Z$=106 isotopes.  This difference of about 3
orders of magnitude between $Z$=106 and $Z$=108 is also
shown in the results of the RMF theory in Fig. 8, where an
enhancement at $N$=166 is seen. The $\alpha$-decay energies
in RMF theory (Fig. 9) are consistent with the experimental
values. Thus, the above experimental data would be
consistent with a strong shell gap at $Z$=106 in the
single-particle spectrum in the RMF theory.

Another shell gap at proton number $Z$=114 is also
predicted in the RMF theory. A similar prediction for
closed proton shell is made by other theories
\cite{Sob.94,MN.94}. However, owing to an enhanced stability
and thus a higher life-time for decay, proton number
$Z$=106 may be preferred in the synthesis of nuclei.  The
shell gaps in neutrons as well as in protons would augment
the stability of nuclei. From this point of view, the
nucleus with $Z$=106 and $N$=166 seems to provide a good
indication for a 'double-magic' nucleus amidst the nuclei
where a significant deformation is prevalent. In other
words, the region $Z$=106 and $N$=166 may provide an island
of extra stability. It is noteworthy that deformation
acquired by nuclei near the above magic region is minimal
as compared to their heavier or lighter counterparts.

In conjunction with the proton number $Z$=114, an
unambiguous shell closure appears at $N$=184 in the RMF
theory, whereby most of the nuclei become spherical. This
shell closure has also been predicted by the the
macroscopic-microscopic calculations of M\"oller and Nix
\cite{MN.94}. In the RMF theory, the combination $Z$=114 and
$N$=184 constitutes another 'double-magic' number. Thus,
RMF theory reinforces the prediction for another region of
extra stability in the domain of superheavy nuclei.

It may be instructive to make a comparison of the RMF
predictions with other theories and models. Some of the
predictions about shell closures as well as on the
existence of an island of extra-stable deformed nuclei in
the RMF theory are consistent with the predictions of the
macroscopic-microscopic approaches. It is noteworthy that
the RMF theory, the FRDM and the Nilsson-Strutinsky
approach \cite{Sob.94}, all indicate strong shell closures
at $Z$=114 and $N$=184.  In other regions, however, various
predictions seem to differ.  Most notable amongst these is
in the middle of the major shell $N$=184, i.e. in the
region of the deformed shell whereby a shell closure
appears at $N$=166 in RMF theory with a prolate deformation
$\beta_2$ = 0.18.  In comparison, the FRDM predicts a
deformed shell closure at $N$=162 and at $N$=164 with a
corresponding $\beta_2$ about 0.22. The prediction of shell
closure in FRDM at $N$=162 is in accord with the prediction
of the Nilsson-Strutinsky approach \cite{Sob.94} where a
deformed shell closure has also been surmised at $N$=162.
The closeness in this prediction of the
macroscopic-microscopic approaches FRDM and
Nilsson-Strutinsky might stem from the similarity in the
calculation of the shell effects.

The FRDM also predicts large gaps in the single-particle
energies at $Z$=104, 106, 108 and 110 corresponding to a
prolate deformation. In contrast, the RMF theory predicts a
strong shell closure at $Z$=106. In the Nilsson-Strutinsky
approach, on the other hand, the deformed shell closure
occurs at $Z$=108.

In conclusion, some of the predictions of the microscopic
RMF theory are consistent with those of the FRDM and
Nilsson-Strutinsky approach whereas some other predictions
of our approach differ from the latter. It is worth
pointing out that the latter models employ a large number
of parameters which are fitted to a large body of data. In
comparison, the RMF theory is based upon a smaller number
of parameters fitted to a limited data. The fact that the
RMF theory using only a few parameters has been able to
describe nuclear properties all over the periodic table
provides a confidence in its predictions.

\section{\sf Acknowledgment}
The authors are thankful to Fedja Ivanyuk for his help in
the calculation of the shell corrections.  One of the
authors (G.A.L) acknowledges support by the European Union
under the contracts HCM-EG/ERB CHBICT-930651 and TMR-EU/ERB
FMBCICT-950216. Partial support from the Bundesministerium
f\"ur Forschung und Technologie under the project 06TM734
(6) is acknowledged.

\newpage 
\leftline{\bf Figure Captions}
\baselineskip = 18pt
\parindent = 7 true cm
\begin{description}

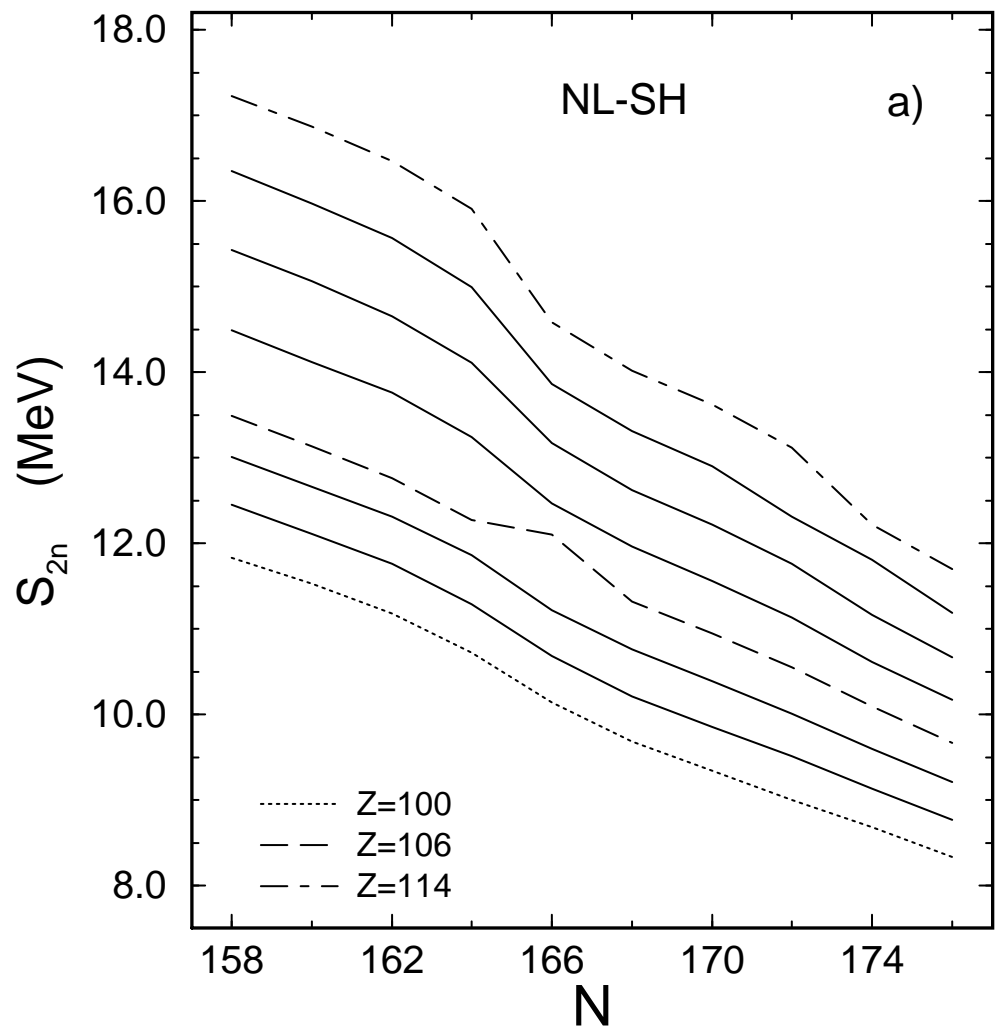
\item[Fig. 1] 
The two-neutron separation energies $S_{2n}$ 
obtained from the RHB calculations with spherical configuration 
for nuclei with atomic numbers $Z$=100 to $Z$=114 for neutron numbers 
(a) $N$=158-176 and (b) $N$=178-190.

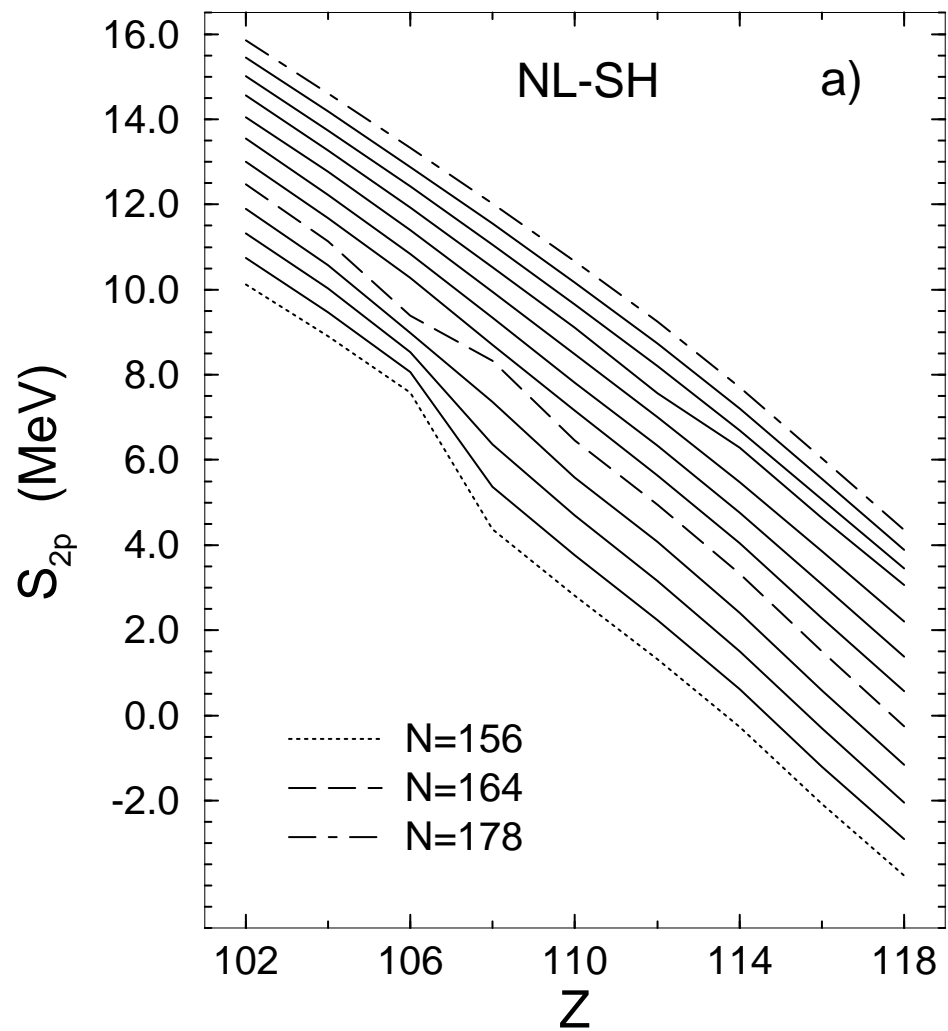
\item[Fig. 2] 
The two-proton separation energies $S_{2p}$ obtained from the 
RHB calculations with spherical configuration 
for nuclei with atomic numbers $Z$=100 to $Z$=114 for neutron numbers 
(a) $N$=158-178 and (b) $N$=180-188.

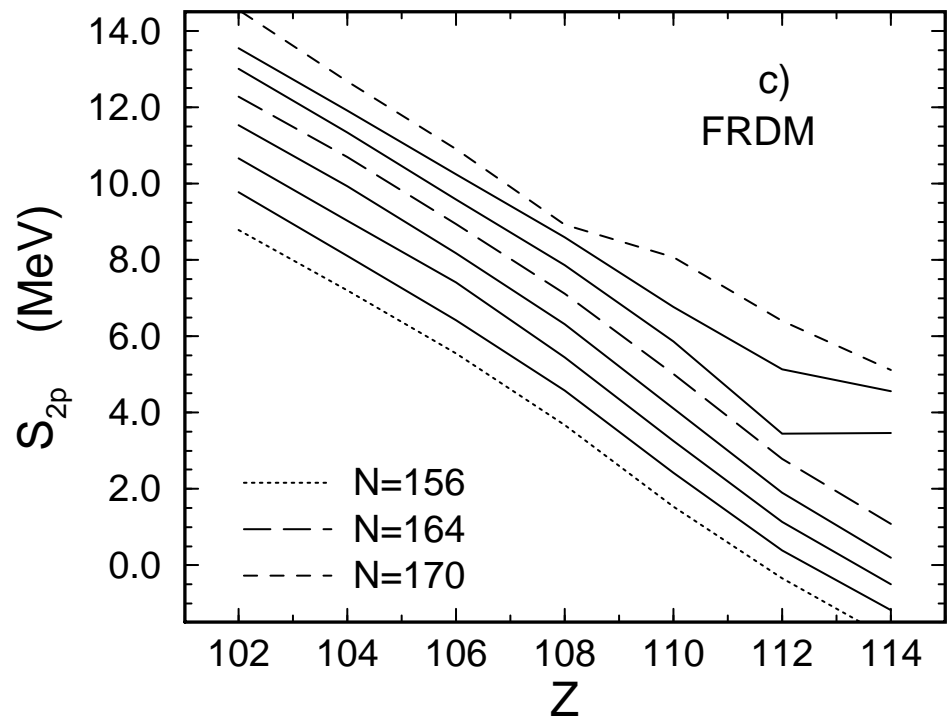
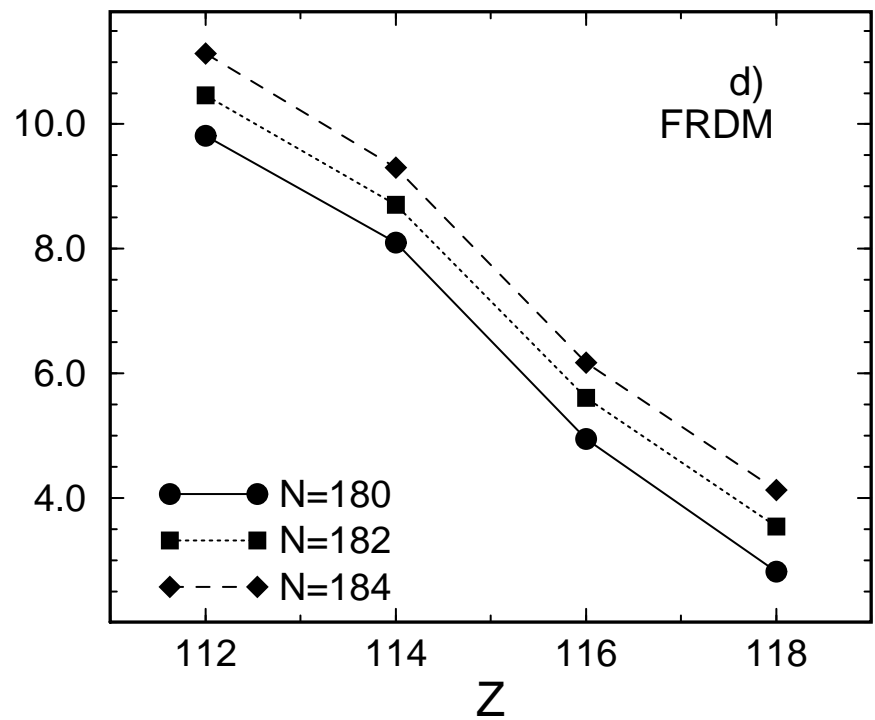
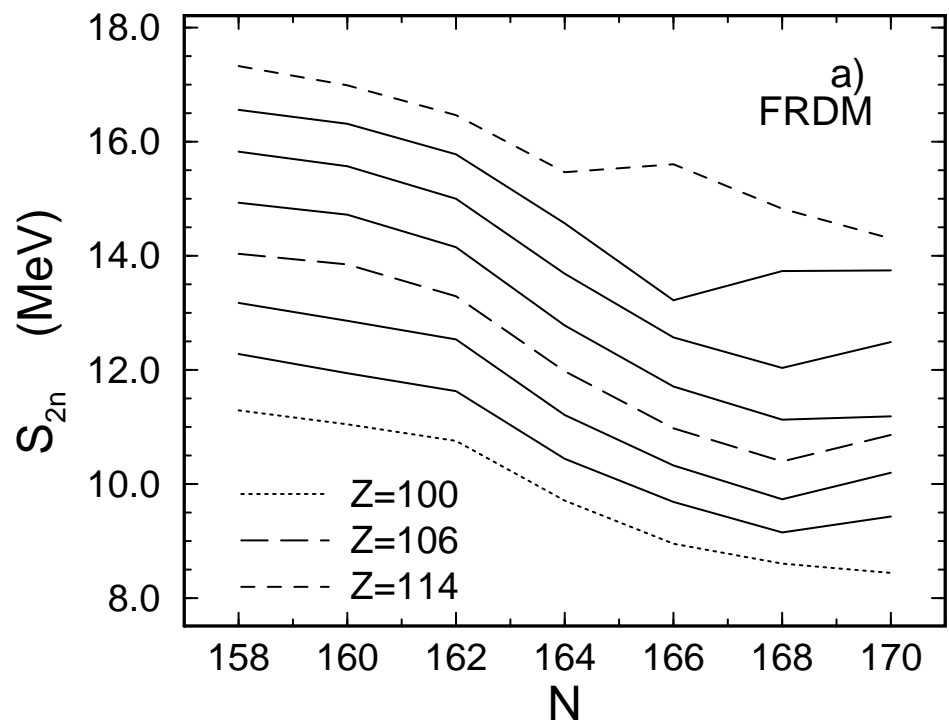
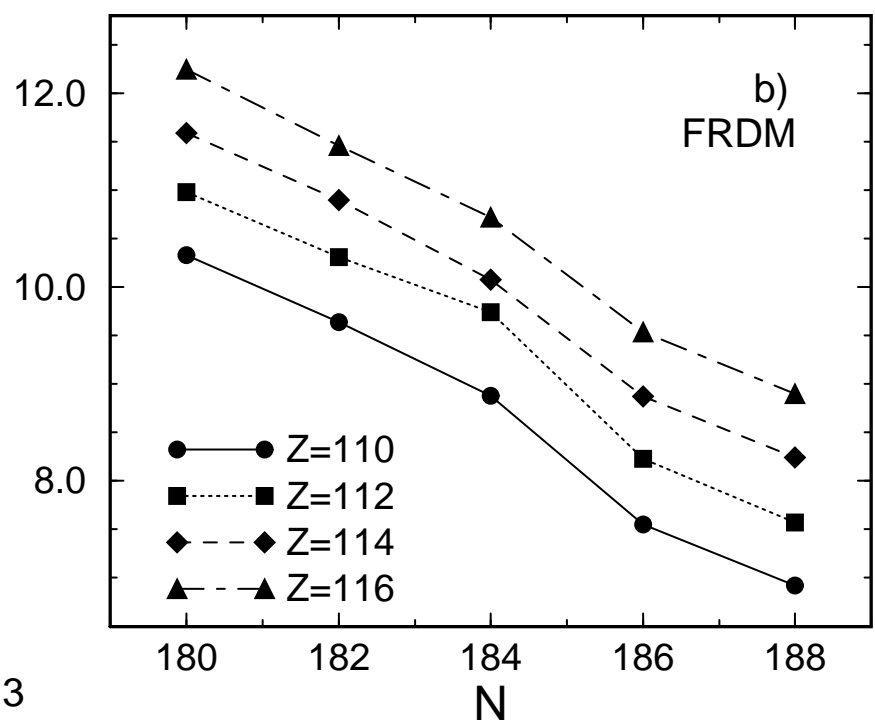
\item[Fig. 3] 
The $S_{2n}$ values from FRDM for nuclei with neutron numbers 
(a) $N$=158-170 and (b) $N$=180-188. The $S_{2p}$ values from FRDM for nuclei
with neutron numbers (c) $N$=156-170 and (d) $N$=180-184, as taken from 
ref. \cite{MNM.95}.

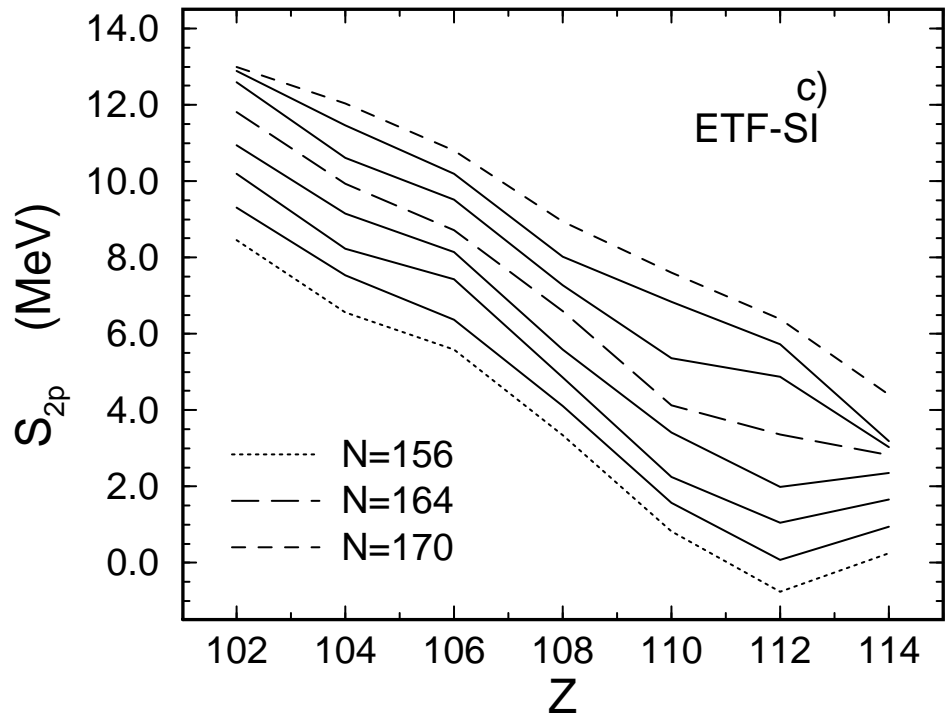
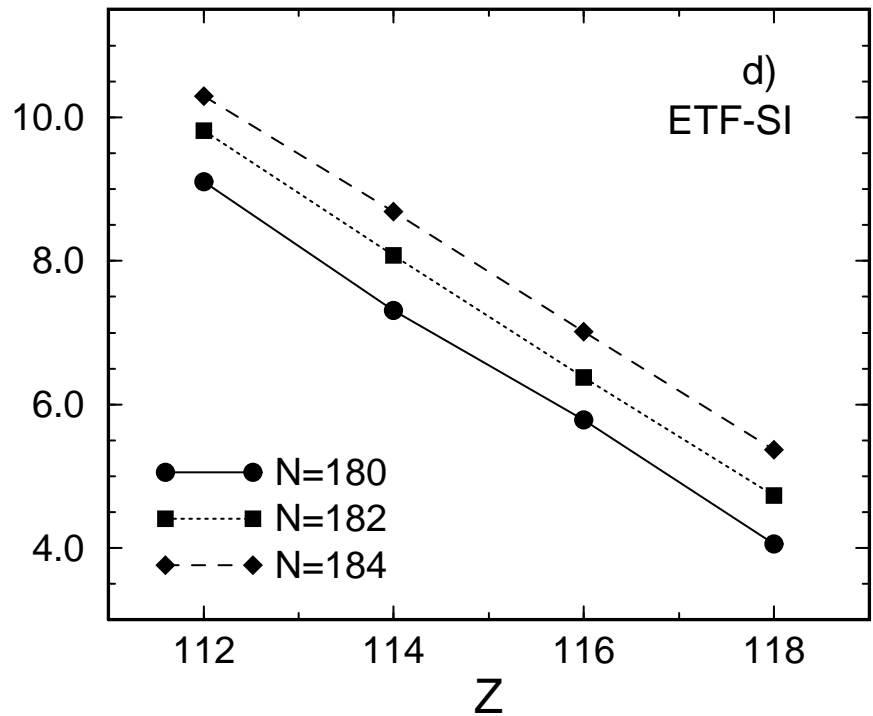
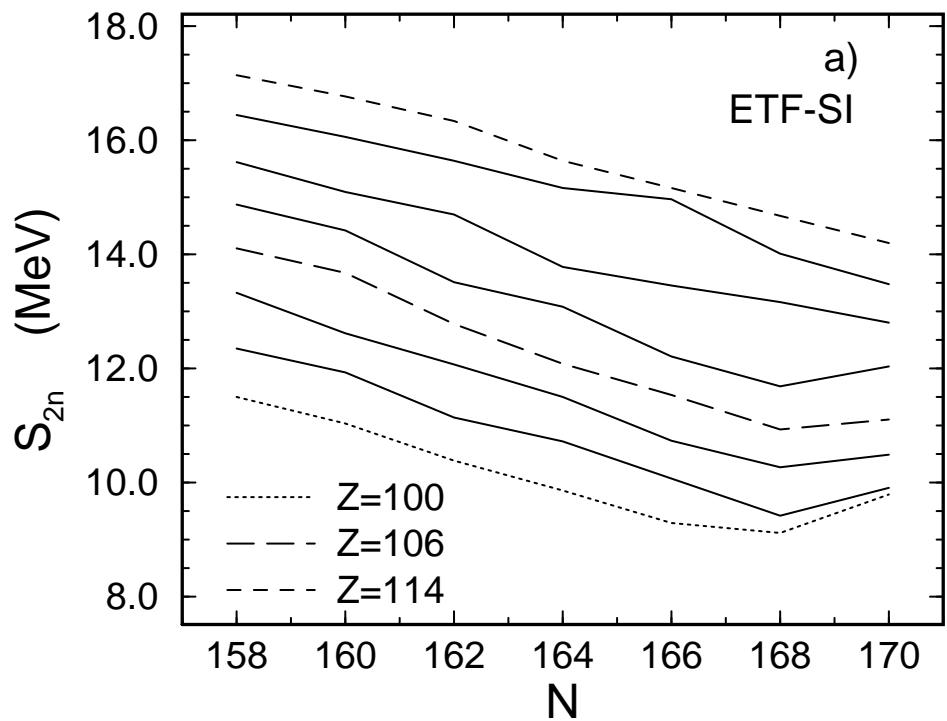
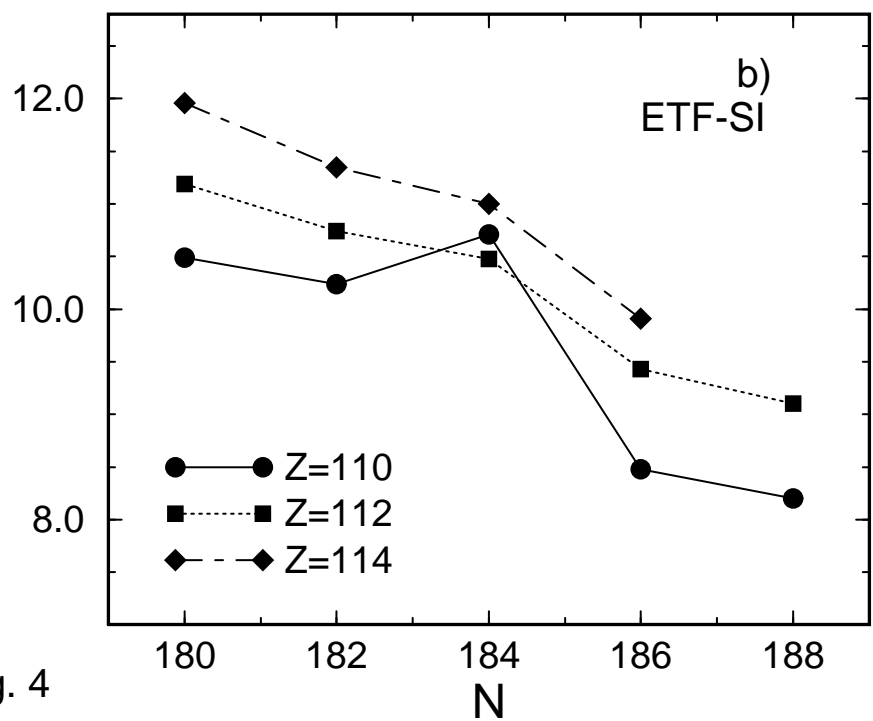
\item[Fig. 4] 
The same as in Fig. 3, but from the ETF-SI as taken from
ref. \cite{APD.92}.

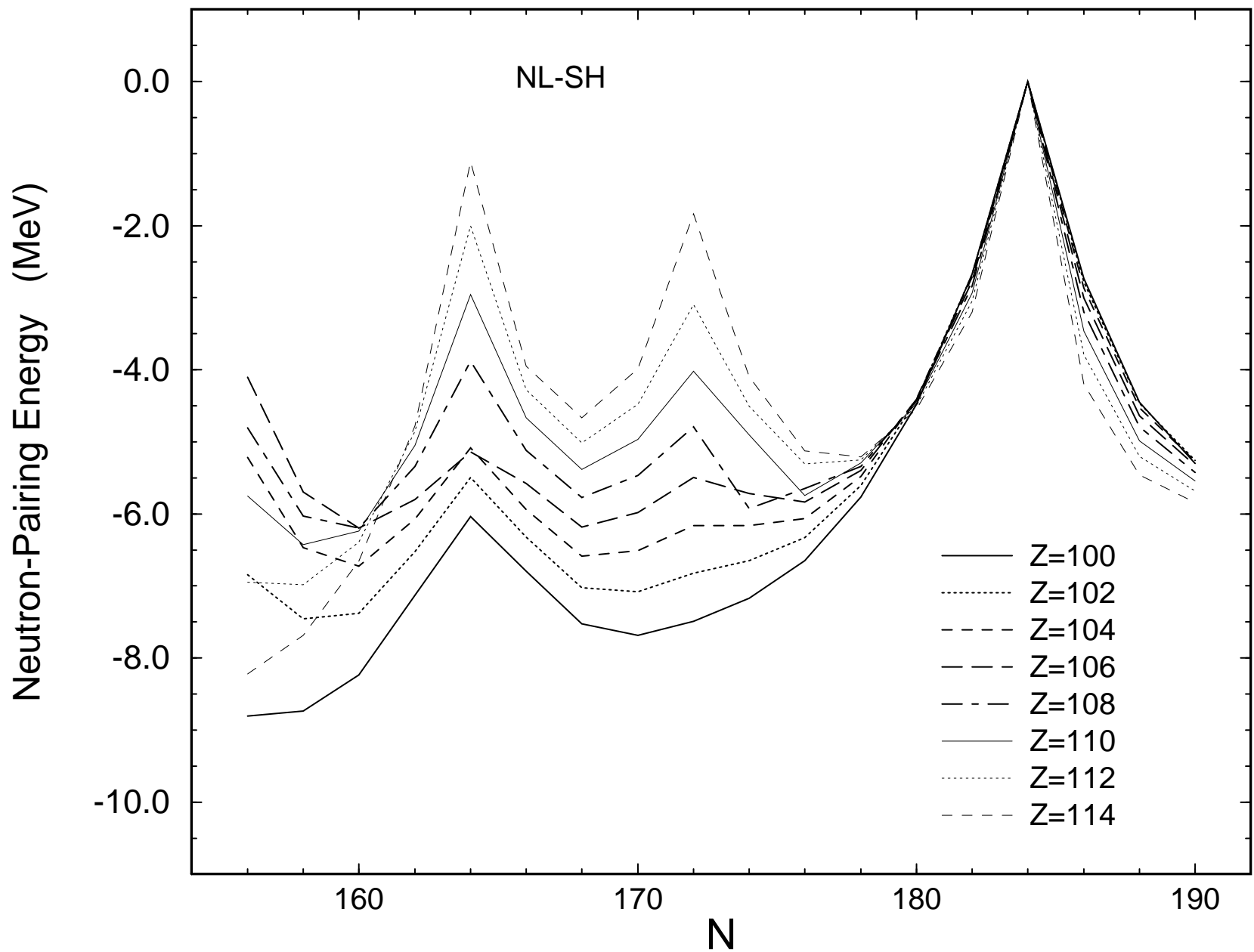
\item[Fig. 5] 
The neutron pairing energy obtained from the RHB
 calculations for spherical configurations. 

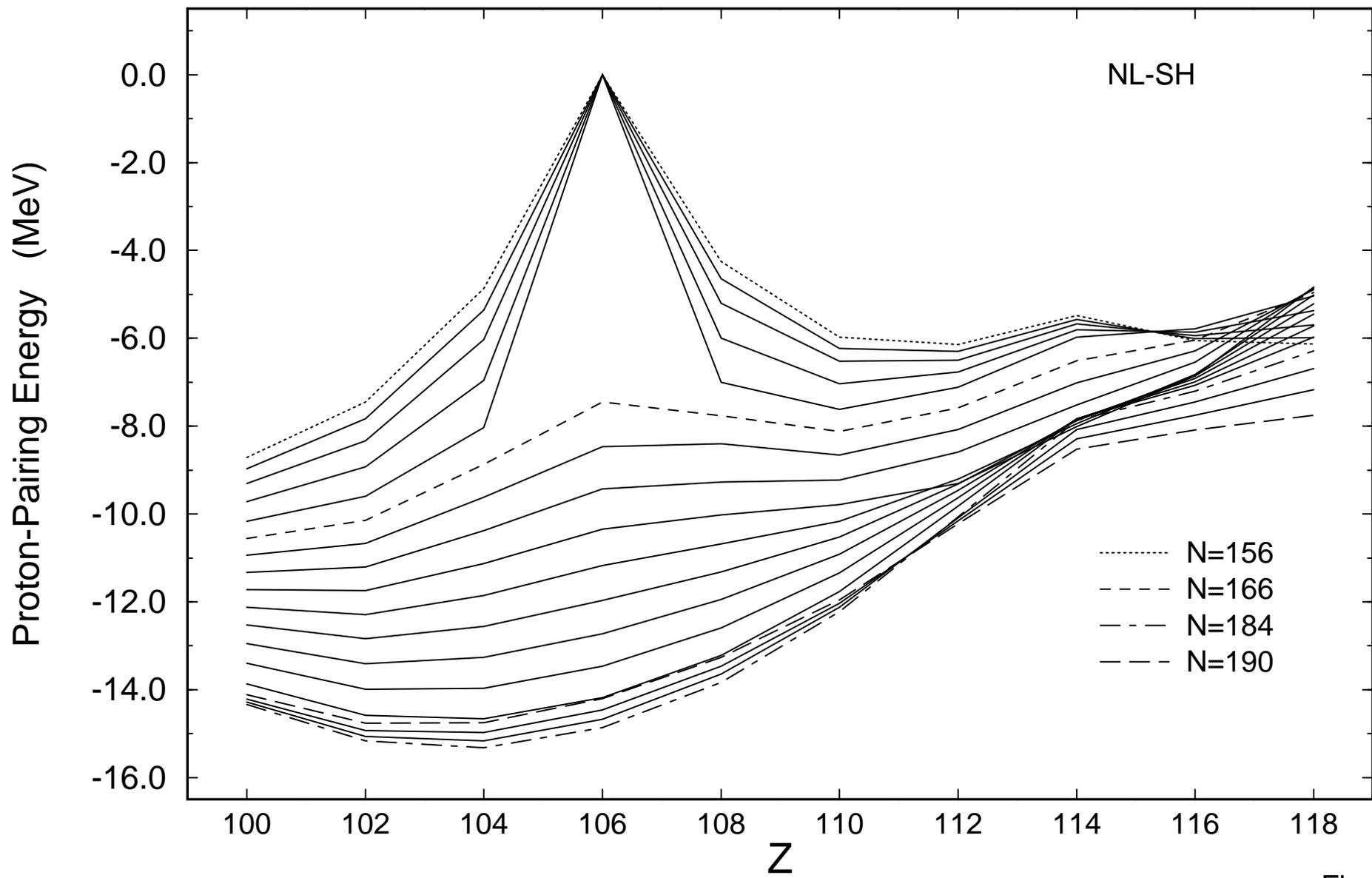
\item[Fig. 6] 
The proton pairing energy obtained from the RHB
calculations for spherical configurations. 

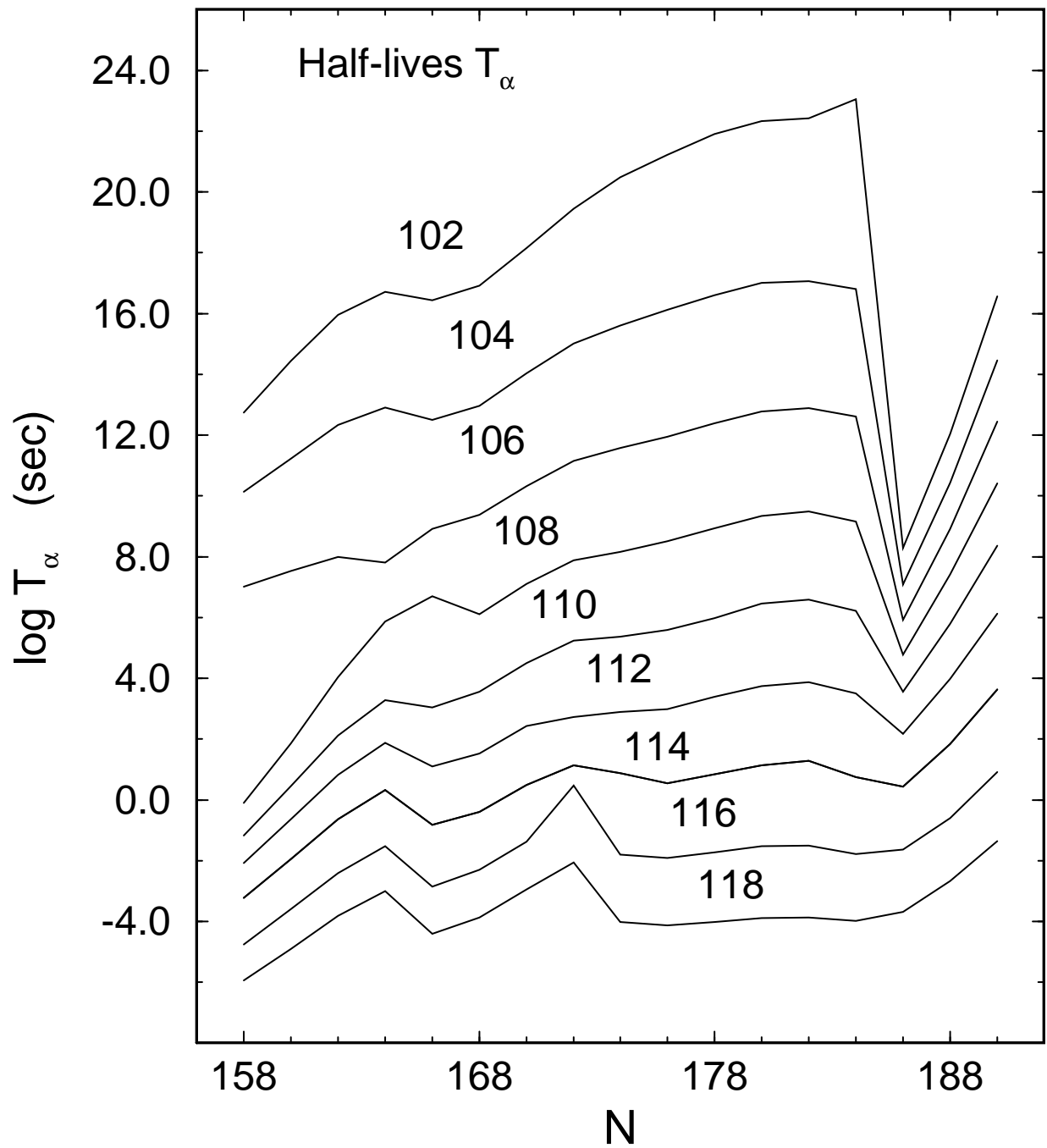
\item[Fig. 7] 
The alpha-decay half life $T_a$ obtained from Viola-Seaborg
systematics using the results of the RHB calculations.

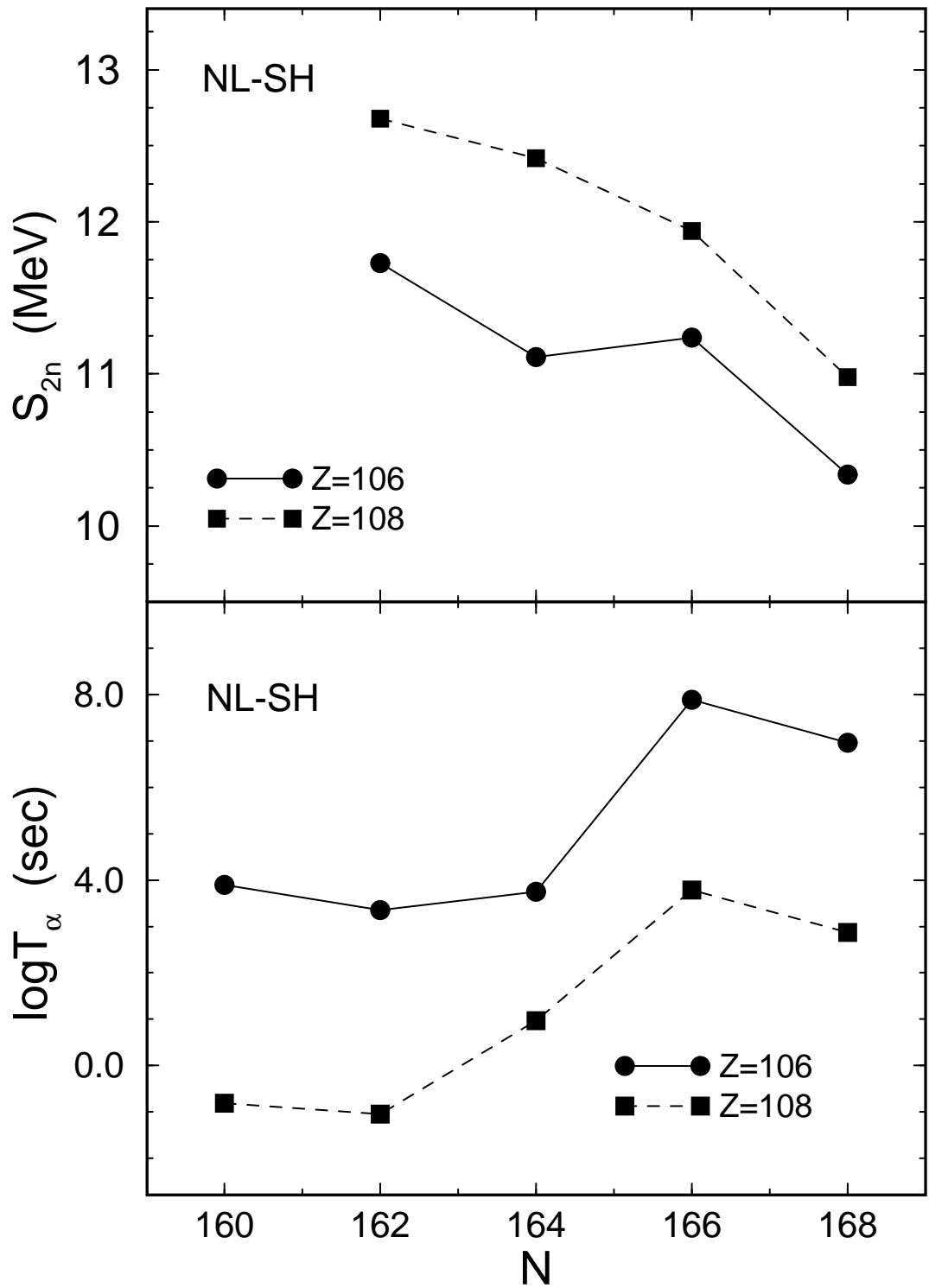
\item[Fig. 8] 
(a) $S_{2n}$ and (b) $T_a$ obtained from the deformed
RMF calculations with 20 oscillator shells.

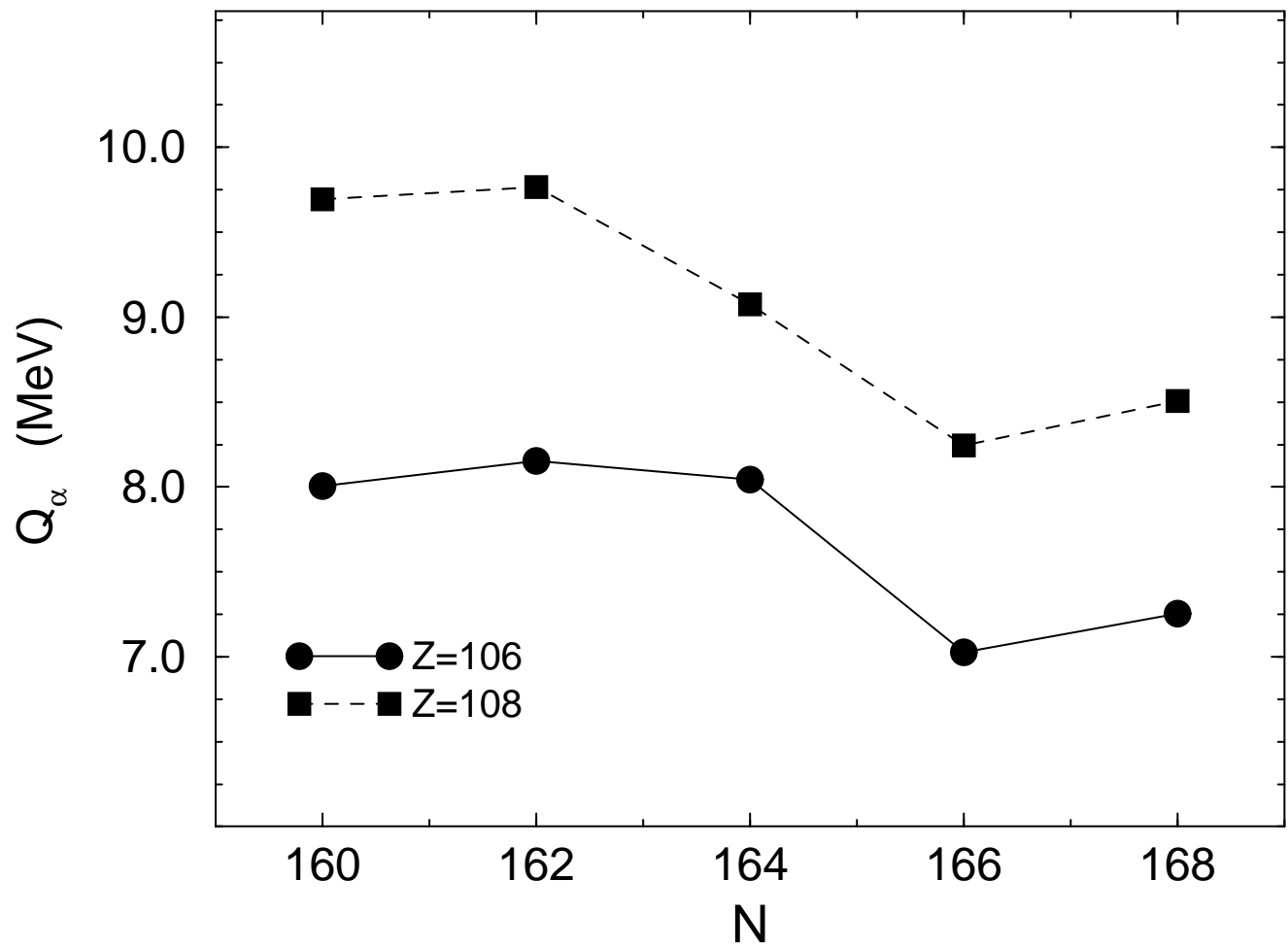
\item[Fig. 9] 
The $Q_\alpha$ values for deformed nuclei about $N$=164
with $Z$=106 and $Z$=108.

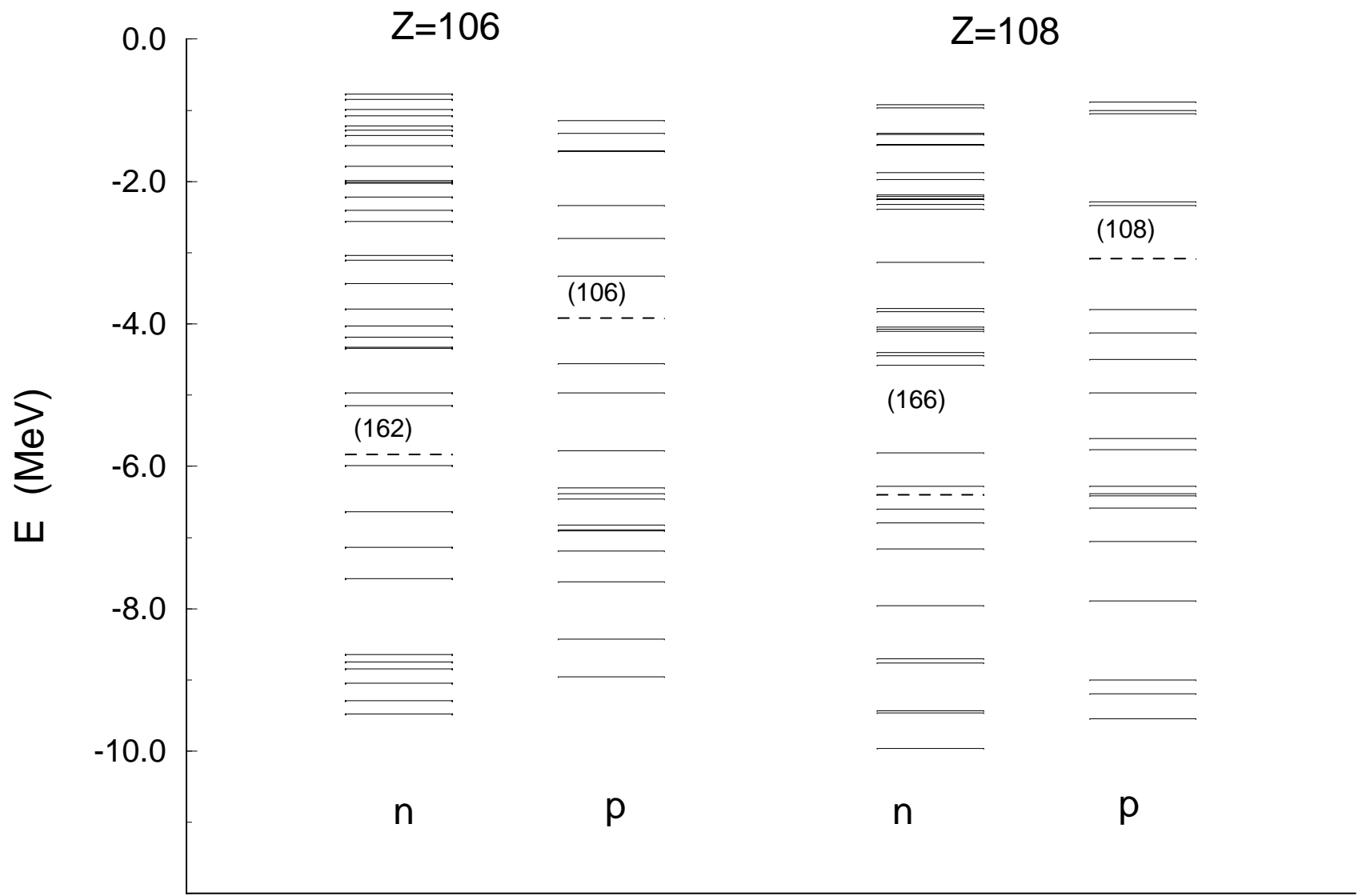
\item[Fig. 10] 
The neutron and proton single-particle energies for
nuclei with $Z$=106 and $Z$=108 with neutron number $N$=162 obtained from
deformed RMF calculations.

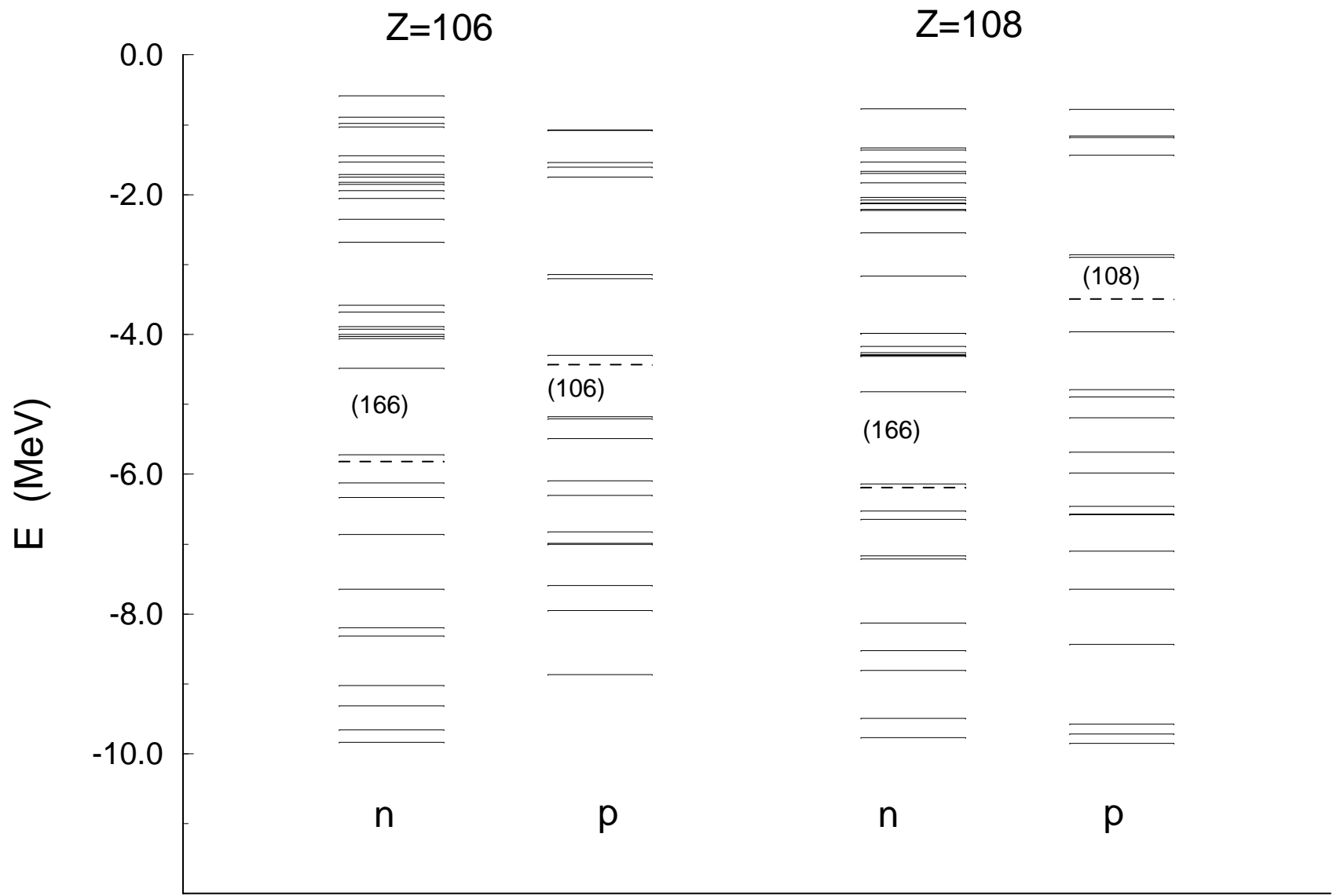
\item[Fig. 11] 
The same as in Fig 10, but for $N$=164.

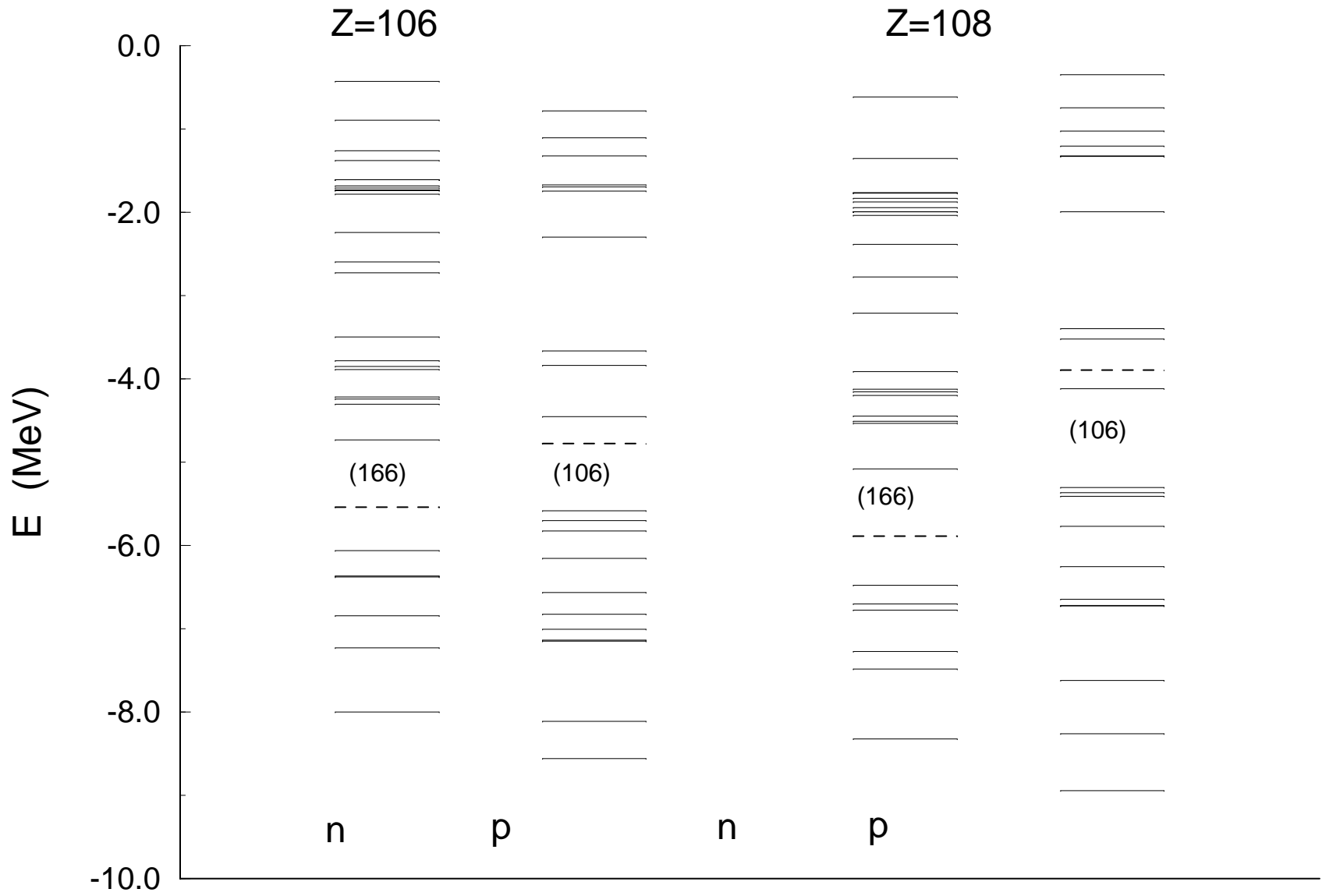
\item[Fig. 12] 
The same as in Fig 10, but for $N$=166.

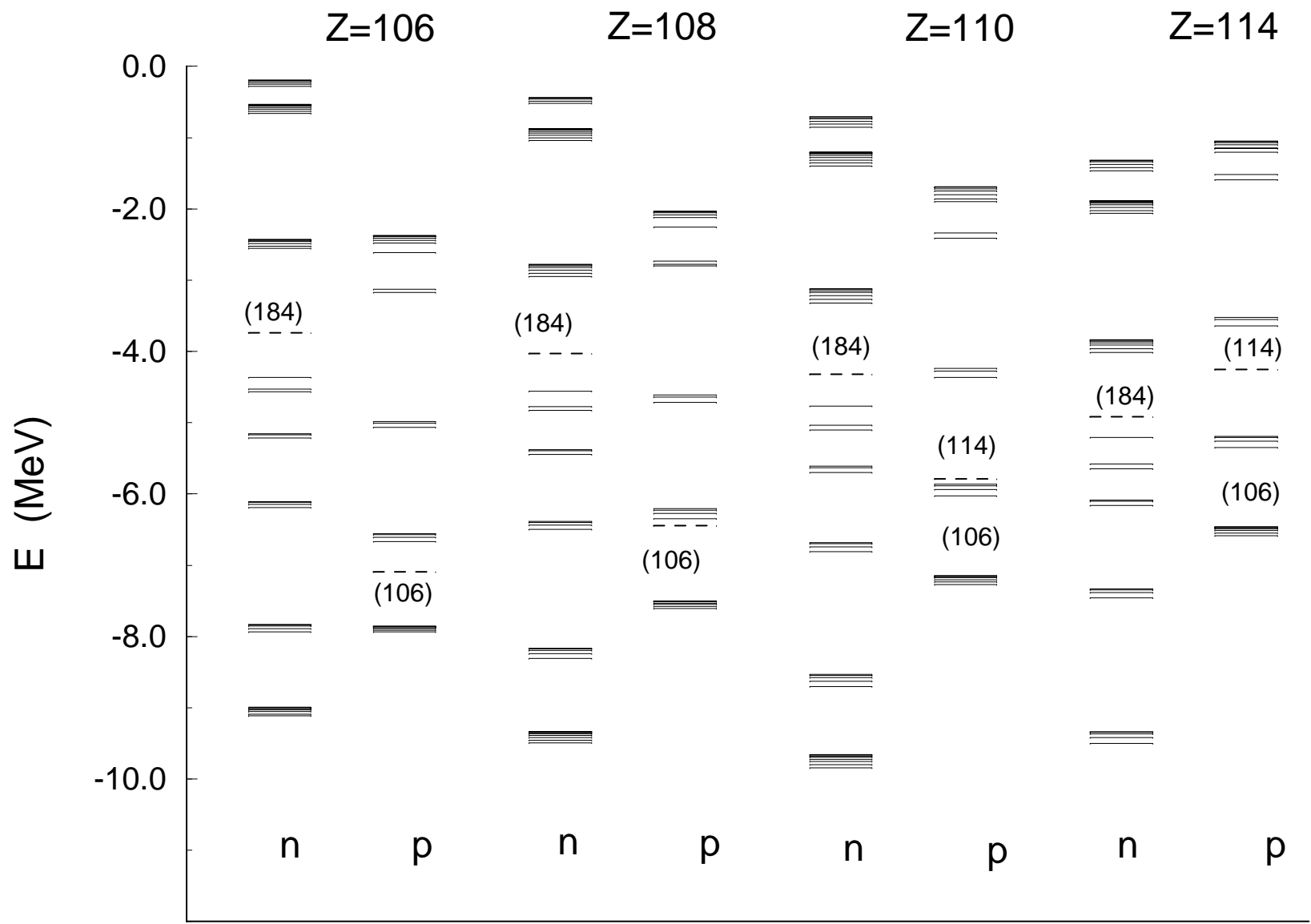
\item[Fig. 13] 
The neutron and proton single-particle energies for
nuclei with $Z$=106, 108, 110 and 114 with neutron number $N$=184.

\end{description}

\newpage
\noindent\begin{table}
\begin{center}
\caption{\sf The relevant parameters of the Gogny interaction D1 used
in the present work. The range is in fermis and the constants W,B,H,M 
are in MeV.}  
\bigskip
\begin{tabular}{ll c c c c c c   l}
\hline\hline
&i  & range  & $W_i$ &  $B_i$  & $H_i$ &  $M_i$ &\\ 
\hline
&1 & 0.7 & -402.4 & -100.0 & -496.2& -23.56& \\
\   \\
&2& 1.2 & -21.3& -11.77& 37.27& -68.81&\\ 
\hline\hline
\end{tabular}
\end{center}
\end{table}

\noindent\begin{table}
\begin{center}
\caption{\sf The parameters of the force NL-SH. All the masses are in MeV,
while $g_{2}$ is in fm$^{-1}$. The other coupling constants are
dimensionless.}
\bigskip
\begin{tabular}{ll c c c c c   l}
\hline\hline
& M = 939.0 &$m_\sigma$ = 526.059 & $m_\omega$ = 783.0 & $m_\rho$
= 763.0 &  &\\
& & & & & & \\
&$g_\sigma$ = 10.444& $g_\omega$ = 12.945& $g_\rho$ = 4.383& $g_2$ = $-$6.9099
&$g_3$ = $-$15.8337&\\ 
\hline\hline
\end{tabular}
\end{center}
\end{table}

\noindent\begin{table}
\begin{center}
\caption{\sf The $\beta_{2}$ and $\beta_{4}$ deformation
parameters calculated in the RMF theory with the NL-SH
force. Values from the mass models FRDM and ETF-SI are
also shown for comparison}
\bigskip
\begin{tabular}{ll c c c c c c c c c  l}
\hline\hline\\
&   &   &    & $\beta_{2}$ &  & &  &  $\beta_{4}$ &    &\\
\hline
& Z & N & RMF & FRDM & ETF-SI & & RMF& FRDM& ETF-SI&\\
\hline
&102& 156& 0.246& 0.228& 0.270& & -0.015& -0.019&  0.000&\\
&   & 158& 0.247& 0.228& 0.250& & -0.029& -0.028& -0.020&\\
\\
&104& 158& 0.251& 0.229& 0.250& & -0.041& -0.037& -0.020&\\
&   & 160& 0.253& 0.220& 0.260& & -0.054& -0.046& -0.050&\\
&   & 162& 0.247& 0.230& 0.260& & -0.060& -0.069& -0.050&\\
&   & 164& 0.197& 0.221& 0.250& & -0.033& -0.072& -0.040&\\ 
&   & 166& 0.179& 0.201& 0.250& & -0.034& -0.067& -0.040&\\
\\
&106& 158& 0.251& 0.229& 0.260& & -0.047& -0.044& -0.050&\\
&   & 160& 0.254& 0.230& 0.260& & -0.060& -0.061& -0.050&\\
&   & 162& 0.248& 0.231& 0.260& & -0.065& -0.078& -0.050&\\
&   & 164& 0.197& 0.221& 0.260& & -0.039& -0.080& -0.050&\\
&   & 166& 0.183& 0.201& 0.250& & -0.042& -0.074& -0.060&\\ 
&   & 168& 0.168& 0.164& 0.230& & -0.043& -0.054& -0.050&\\
\\
&108& 160& 0.236& 0.230& 0.260& & -0.052& -0.069& -0.050&\\
&   & 162& 0.211& 0.231& 0.260& & -0.042& -0.086& -0.080&\\
&   & 164& 0.198& 0.222& 0.250& & -0.045& -0.089& -0.070&\\
&   & 166& 0.185& 0.212& 0.250& & -0.049& -0.091& -0.070&\\ 
&   & 168& 0.173& 0.164& 0.410& & -0.053& -0.063&  0.070&\\
\\
&110& 166& 0.182& 0.212& 0.410& & -0.058& -0.091&  0.070&\\ 
\\
&112& 166& 0.180& 0.164& 0.430& & -0.065& -0.063&  0.070&\\
&   & 168& 0.173& 0.080& 0.430& & -0.071& -0.006&  0.070&\\
\hline\hline
\end{tabular}
\end{center}
\end{table}

\newpage
\noindent\begin{table}
\begin{center}
\caption{\sf The binding energies (in MeV) for some
superheavy nuclei with $N$=184 obtained in the deformed RMF
calculations with the interaction NL-SH.  Predictions from
the mass models FRDM and ETF-SI are also shown for
comparison. The quadrupole deformation $\beta_2$ obtained
in the RMF calculations along-with the predictions from FRDM
and ETF-SI.  The RMF theory predicts a spherical shape for
these nuclei.}
\bigskip
\begin{tabular}{ll c c c c c c c l}
\hline\hline
& & & B.E& & & & $\beta_{2}$& & \\
\hline    
&A& NL-SH &FRDM & ETF-SI& &NL-SH &FRDM&  ETF-SI & \\
\hline 
&$^{290}$106&$-$2078.65&$-$2074.10&$-$2078.25& & $-$0.003& 0.000& ~0.000&\\
&$^{292}$108&$-$2092.15&$-$2088.56&$-$2091.60& &$-$0.004& 0.000& ~0.000&\\
&$^{294}$110&$-$2104.38&$-$2101.67&$-$2103.60& &$-$0.005& 0.000&$-$0.010&\\  
&$^{298}$114&$-$2125.00&$-$2123.30&$-$2122.86& & $-$0.005& 0.000&$-$0.010&\\ 
\hline\hline
\end{tabular}
\end{center}
\end{table}

\noindent
\begin{table}
\begin{center}
\caption{\sf The shell corrections calculated in the RMF theory
 for a number of heavy nuclei. The effect due to pairing has been 
included in obtaining the shell energies.}
\bigskip
\begin{tabular}{ll c c c c c c c c    l}
\hline\hline\\
&Z/N & 154 & 156 & 158  & 160 & 162 & 164 & 166 & 168& \\ 
\hline
&104&-6.22&-6.32&-6.47&-6.79&-6.16&-3.76&-3.35&-2.60&\\
&106&-6.44&-6.97&-7.53&-8.05&-7.27&-4.88&-4.58&-3.64&\\
&108&-6.54&-7.12&-7.32&-7.12&-6.28&-5.74&-5.43&-4.40&\\
&110&-6.47&-6.20&-6.39&-6.21&-5.86&-5.85&-6.12&-5.32&\\
&112&-4.43&-5.40&-5.67&-5.60&-5.73&-6.37&-7.09&-6.49&\\
&114&-3.86&-4.71&-4.87&-4.79&-4.97&-5.91&-6.79&-6.66&\\
&116&-3.41&-4.21&-4.23&-4.00&-3.96&-4.84&-5.78&-5.75&\\
\hline\hline
\hline\hline
\end{tabular}
\end{center}
\end{table}
\vfill

\newpage
\baselineskip = 14pt
\begin{thebibliography}{999}
\bibitem{Nix.77} J.R. Nix, Physics Today  {1972} p. 30.

\bibitem{Sob.94} A. Sobiczewski,  Sov. J. Part. Nucl. {\bf 25} (1994) 119. 

\bibitem{MN.94} P. M\"oller, J.R. Nix,  J. Phys. {\bf G20} (1994) 1681.

\bibitem{Amb.85} P. Ambruster, Ann. Rev. Nucl. Part. Sci.
        {\bf 35} (1985) 135.

\bibitem{OL.85} Yu. Ts. Oganessian, and Yu. A. Lazarev, in
        {\it Treaties on Heavy-Ion Science}, Vol. 4, edited
        by D.A. Bromley (Plenum Press, N.Y., 1985) p. 3.

\bibitem{Mun.88} G. M\"unzenberg, Rep. Prog. Phys. {\bf 51} (1988) 57.

\bibitem{Amb.94} P. Ambruster, in Proc. of Inter. Conf. on
        Nuclear Shapes and Nuclear Structure at Low Excitation
        Energies, Antibes (France), (eds.) M. Vergnes, D. Goutte,
        P.H. Heenen and J. Sauvage, Editions Frontieres,
        (1994) p. 365.

\bibitem{Oga.95} Yu. Ts. Oganessian, Nucl. Phys. {\bf A583} (1995) 823.

\bibitem{Hof.95} S. Hofmann et al., Z. Phys. {\bf A350} (1995) 277.

\bibitem{Hof.96} S. Hofmann et al., Z. Phys. {\bf A354} (1996) 229.

\bibitem{Laz.94} Yu. A. Lazarev et al., Phys. Rev. Lett. {\bf 73} (1994) 624.

\bibitem{Laz.95} Yu. A. Lazarev et al., Phys. Rev. Lett. {\bf 75} (1995) 1903.

\bibitem{Laz.96} Yu. A. Lazarev et al,. Dubna Report No. E7-95-552 (submitted
   to Phys. Rev. C).

\bibitem{pat91} Z. Patyk and A. Sobiczewski,  Nucl. Phys. {\bf A533} 
(1991) 132.

\bibitem{pat91b} Z. Patyk and A. Sobiczewski, Phys. Lett. {\bf 256B} 
(1991) 307. 

\bibitem{bei74} M. Beiner, H. Flocard and M. Veneroni, Phys. Scr. {\bf 10A}
(1974) 84.

\bibitem{quent78} P. Quentin and H. Flocard, Ann. Rev. Nucl. Part. Sci.
{\bf 28} (1978) 523.

\bibitem{GRT.90}Y.K. Gambhir, P. Ring, and A. Thimet, 
    Ann. Phys. (N.Y.) {\bf 198} (1990) 132.

\bibitem{Dob.96} J. Dobaczewski, private communication to P. Ring                     
\bibitem{rein89} P.G. Reinhard, Rep. Prog. Phys. {\bf 55} (1989) 439.

\bibitem{SNR.93}M.M. Sharma, M.A. Nagarajan, and P. Ring,
    Phys. Lett. {\bf B312} (1993) 377

\bibitem{LS.95}G.A. Lalazissis and M.M. Sharma, 
    Nucl. Phys. {\bf A586} (1995) 201.

\bibitem{LSR.95} G.A. Lalazissis, M.M. Sharma and P. Ring, 
        Nucl. Phys. {\bf A597} (1996) 35.

\bibitem{SLH.94} M.M. Sharma, G.A. Lalazissis, W. Hillebrandt, and P. Ring,
     Phys. Rev. Lett. {\bf 72} (1994) 1431.

\bibitem{SLR.93}M.M. Sharma, G.A. Lalazissis, and P. Ring,
    Phys. Lett. {\bf B317} (1993) 9.

\bibitem{KR.91}H. Kucharek and P. Ring Z. Phys. {\bf A339} (1991) 23.

\bibitem{SW.86}B.D. Serot and J.D. Walecka, 
    Adv. Nucl. Phys. {\bf 16} (1986) 1.

\bibitem{Ser.92}B.D. Serot, Rep. Prog. Phys. {\bf 55} (1992) 1855.

\bibitem{BB.77}J. Boguta and A.R. Bodmer, Nucl. Phys. {\bf A292} (1977) 413.

\bibitem{DG.80} J. Decharge and D. Gogny, Pfys. Rev. {\bf C 21} (1980) 1568.

\bibitem{Tal} J.D.  Talman, Nucl. Phys. {\bf A141} (1970) 273. 

\bibitem{MN.92}P. M\"oller, J.R. Nix,  Nucl. Phys. {\bf A536} (1992) 20.

\bibitem{AW.93}G. Audi and A.H. Wapstra, Nucl. Phys. {\bf A565} (1993) 1.

\bibitem{MNM.95}P. M\"oller, J.R. Nix, W.D. Myers, and W.J. Swiatecki, 
    Atomic Data and Nuclear Data Tables 59 (1995) 185.

\bibitem{APD.92}Y. Aboussir, J.M. Pearson, A.K. Dutta, and F. Tondeur,
     Atomic Data and Nuclear Data Tables 61 (1995) 127.

\bibitem{viola} V.E. Viola Jr. and G.T. Seaborg, J. Inorg. Nucl. Chem.
{\bf 28} (1966) 741.

\bibitem{sobi89} A. Sobiczewski,  Z. Patyk and S. Cwiok,  Phys. Lett. 
{\bf 224B} (1989) 1.

\bibitem{LQ.82}J. Libert and P. Quentin, Phys. Rev. {\bf C25} (1982) 571.

\bibitem{BDJ.72} M. Brack, J. Damgaard, A.S. Jensen, H.C. Pauli, 
        V.M. Strutinsky, C.Y.Wong; Rev. Mod. Phys. {\bf 44} (1972) 320.


\end{thebibliography}
\end{document}